\documentclass[11pt,dvips]{article}

\usepackage{epsfig}
\usepackage{graphicx,float}
\usepackage{array}
\usepackage{tabularx}
\usepackage{indentfirst}
\usepackage{amsmath,amssymb,bbold}
\usepackage{url}
\usepackage{rotating}
\usepackage{color}
\usepackage{hhline}
\usepackage{hyperref}
\usepackage{slashed}
\usepackage{tikz}
\usepackage{makecell}
\usepackage{pgfplots}
\usepackage{empheq}
\usepackage{physics}
\usepackage{cite}

\newcommand{\s}{\\ \vspace*{-3.5mm}}

\renewcommand{\thefootnote}{\arabic{footnote}}

\begin{document}


\begin{flushright}
\today \\
\end{flushright}

\vskip 0.5cm

\begin{center}
{\Large \bf Weaving the covariant three-point vertices efficiently}\\[1.0cm]
{Seong Youl Choi\footnote{sychoi@jbnu.ac.kr} and
 Jae Hoon Jeong\footnote{jaehoonjeong229@gmail.com}} \\[0.5cm]
{\it  Department of Physics and RIPC,
      Jeonbuk National University, Jeonju 54896, Korea}
\end{center}

\vskip 0.5cm

\begin{abstract}
\noindent
An efficient algorithm is developed for compactly weaving all the Lorentz
covariant three-point vertices in relation to the decay of a massive
particle $X$ of mass $m_X$ and spin $J$ into two particles $
M_{1,2}$ with equal mass $m$ and spin $s$. The closely-related
equivalence between the helicity formalism and the covariant formulation
is utilized so as to identify the basic building blocks
for constructing the covariant three-point vertex corresponding to
each helicity combination explicitly. The massless case with $m=0$ is
worked out straightforwardly and the (anti)symmetrization of the
three-point vertex required by spin statistics of identical particles
is made systematically. It is shown that the off-shell electromagnetic 
photon coupling to the states $M_1$ and $M_2$ can be accommodated 
in this framework. The power of the algorithm is demonstrated 
with a few typical examples with specific $J$ and $s$ values.
\end{abstract}



\vskip 0.8cm

\setcounter{footnote}{0}
\renewcommand{\thefootnote}{\alph{footnote}}

\section{Introduction}
\label{sec:introduction}

The Standard Model (SM)~\cite{Glashow:1961tr,Weinberg:1967tq,Salam:1968rm,Fritzsch:1973pi}
of particle physics has been firmly established by the discovery of
the spin-0 resonance of about 125\, GeV mass at the Large Hadron Collider
(LHC) at CERN~\cite{Aad:2012tfa,Chatrchyan:2012ufa}. However,
even though a lot of high-energy experiments have searched for new phenomena
beyond the SM (BSM) and they have tested the SM with great precision
for decades, none of any BSM particles and phenomena has been observed
so far at the TeV scale (see Ref.~\cite{Zyla:2020zbs} for a comprehensive
summary of hypothetical particles and concepts). In this present
situation considered to be unnatural, one meaningful strategy that can
be taken in new physics searches is to keep our theoretical studies
as model-independent as possible and to search for new particles
with even more exotic characteristics including {\it spins higher
than unity}. \s

Recently, the theory and phenomenology of high-spin particles~\cite{Shklyar:2009cx,Bergshoeff:2016soe,Jafarzade:2021vhh,
Babichev:2016hir,Babichev:2016bxi,Marzola:2017lbt,
Criado:2020jkp,Falkowski:2020fsu,Gondolo:2020wge,Gondolo:2021fqo,
Criado:2021itq} have drawn considerable interest.
Various high-spin particles, although composite, exist in hadron
physics (see Ref.~\cite{Zyla:2020zbs} for several high-spin
hadrons) so that the solid theoretical calculations of all the
rates and distributions involving such high-spin states are required
for correctly
interpreting all the relevant experimental
results~\cite{Shklyar:2009cx,Bergshoeff:2016soe,Jafarzade:2021vhh}.
A popular spin-3/2 particle is the gravitino appearing
as the supersymmetric partner of the massless spin-2 graviton
in supergravity~\cite{Nath:1975nj,Volkov:1973jd,Freedman:1976xh,
Deser:1976eh,Fayet:1977vd}.
The discovery of gravitational
waves~\cite{Abbott:2016blz,TheLIGOScientific:2017qsa,
LIGOScientific:2018mvr} strongly indicates the existence of
massless spin-2 gravitons at the quantum level.
The massive
spin-2 particles as the Kaluza-Klein (KK) excitations of the massless
graviton have been studied in the context of extra
dimensional scenarios~\cite{Antoniadis:1998ig,ArkaniHamed:1998rs,
Randall:1999ee}.
In addition, whether the dark matter (DM) of the Universe is formed
with high-spin particles has been addressed in various recent
works~\cite{Babichev:2016hir,Babichev:2016bxi,Marzola:2017lbt,
Criado:2020jkp,Falkowski:2020fsu,Gondolo:2020wge,
Gondolo:2021fqo}. The relic density of the high-spin DM particles
and their low-energy interactions with the SM particles have to be
evaluated precisely for checking the plausibility of their indirect
and direct observations. {\it For studying all of these theoretical and phenomenological aspects, it is crucial to systematically investigate
all the allowed effective interactions of high-spin particles as well
as the SM spin-$0$, $1/2$ and $1$ particles in a model-independent way.}  \s

In the present work, we focus on developing an efficient algorithm
for compactly weaving all the three-point vertices consistent
with {\it Lorentz invariance and locality}.\footnote{If necessary,
any other symmetry principles like local gauge invariance and/or
various discrete symmetries
could be invoked.} Specifically, we consider the decay
of a massive particle of mass $m_X$ and spin $J$ into two massive
particles, $M_1$ and $M_2$, of equal mass $m$ and spin $s$. This
study is a natural extension of the previous work~\cite{Choi:2021ewa}
having dealt with the massless ($m=0$) case with the identical particle (IP)
condition imposed, and an intermediate bridge
to the general case where all the three particles have different
masses and spins.
We adopt the conventional description of the integer and half-integer
wave tensors~\cite{Behrends:1957rup,Auvil:1966eao,Caudrey:1968vih,
Scadron:1968zz,Chung:1997jn,Huang:2003ym,Miyamoto:2011aa} and we effectively
utilize the closely-related equivalence between the helicity formalism in
the Jacob-Wick (JW) convention~\cite{Jacob:1959at,Haber:1994pe}
and the standard covariant formulation. Their one-to-one
correspondence enables us to identify every basic building block
for constructing the covariant three-point vertex corresponding to
each helicity combination explicitly.\footnote{Another
convenient procedure for describing the three-point vertex of massive
particles of any spin is to use a spinor formalism developed in Ref.~\cite{Arkani-Hamed:2017jhn}.} we show that the massless ($m=0$)
case treated previously~\cite{Choi:2021ewa} can be worked out
straightforwardly and the (anti)symmetrization of the vertex required
by spin statistics of identical particles~\cite{Boudjema:1990st} can
be made systematically.
The extension of the algorithm to the case where all the three
particles have different masses and spins is briefly touched upon.\s

This paper is organized as follows.
In Section~\ref{sec:characterization_in_the_helicity_formalism},
we discuss the key aspects of the two-body decay process
$X\to M_1M_2$ in the helicity formalism which allows us to treat
any massive and massless particles on an equal footing.
Section~\ref{sec:spin-j_spin-s_wave_tensors} is devoted to
explicitly deriving and characterizing the integer and half-integer
spin-$s$ wave tensors and the spin-$J$ wave tensor based on the
conventional combination of spin-1 wave vectors and spin-1/2 spinors.
In particular, the wave vectors and spinors in the $X$ rest frame ($X$RF)
are presented explicitly in the JW convention.
Utilizing the close inter-relationship between the helicity
formalism and the covariant formulation we derive all the covariant
basic and composite three-point operators
both in the bosonic and fermionic case
in Section~\ref{sec:basic_covariant_three_point_vertices}.
In Section~\ref{sec:weaving_covariant_three_point_vertices}, the
composite operators are shown to enable us to explicitly write down
all the helicity-specific three-point vertices through which the
covariant three-point vertex for any spin $J$ and spin $s$ can be
weaved efficiently both in the massive and massless cases.
In addition, we show that the relation valid in the case with
two identical particles in the final state is systematically
and straightforwardly derived and that the off-shell electromagnetic
photon coupling to the states $M_1$ and $M_2$ can be
accommodated in this framework.
In order to demonstrate the power of the developed algorithm for
constructing the covariant three-point vertices, we work out
in detail several examples with specific $J$ and $s$ values, which are
expected to be useful for various phenomenological investigations, in Section~\ref{sec:various_specific_examples}. Finally,
we summarize our findings and conclude the present work by mentioning a few
topics under study in Section~\ref{sec:conclusions}.\s

\setcounter{equation}{0}

\section{Characterization in the helicity formalism}
\label{sec:characterization_in_the_helicity_formalism}

\begin{figure}[htb]
\begin{center}
\includegraphics[width=10.cm,height=6.cm]{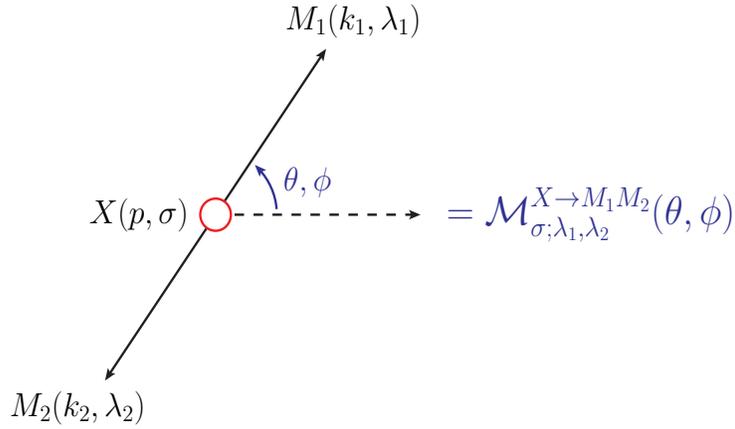}
\caption{\it Kinematic configuration for the helicity amplitude
             ${\cal M}^{X\to M_1 M_2}_{\sigma;\,\lambda_1,\lambda_2}
             (\theta,\phi)$ of the two-body decay $X\to M_1M_2$ of a
             massive particle $X$ into two massive particles, $M_1$ and
             $M_2$, in the $X$RF.
             The notations, $\{p, k_{1,2}\}$ and $\{\sigma, \lambda_{1,2}\}$,
             are the momenta and helicities of the decaying particle $X$ and
             two particles, $M_1$ and $M_2$, respectively. The polar and
             azimuthal angles, $\theta$ and $\phi$, are defined with respect
             to an appropriately chosen coordinate system.
}
\label{fig:kinematic_configuration_xm1m2_xrf}
\end{center}
\end{figure}

The helicity formalism~\cite{Jacob:1959at,Haber:1994pe} allows us to
efficiently describe the two-body decay of a spin-$J$ particle $X$ of
mass $m_X$ into two massive particles, $M_1$ and $M_2$, with equal mass
$m$ and spin $s$. For the sake of a transparent and straightforward
analytic analysis, we describe the two-body decay $X\to M_1M_2$ in the
$X$RF
\begin{eqnarray}
X(p,\sigma)
   \ \ \rightarrow\ \
M_1(k_1,\lambda_1)\, +\, M_2(k_2,\lambda_2)\,,
\label{eq:x_m1m2_decay}
\end{eqnarray}
in terms of the momenta, $\{p, k_1, k_2\}$, and
helicities, $\{\sigma, \lambda_1, \lambda_2\}$, of the particles,
as depicted in Figure~\ref{fig:kinematic_configuration_xm1m2_xrf}.\s

\vskip 0.5cm

The helicity amplitude of the decay $X\to M_1M_2$ is decomposed in terms
of the polar and azimuthal angles, $\theta$ and $\phi$, defining the
momentum direction of the particle $M_1$ in a fixed coordinate system
as
\begin{eqnarray}
  {\cal M}^{X\to MM}_{\sigma;\lambda_1,\lambda_2}(\theta,\phi)
\,\, =\,\, {\cal C}^J_{\lambda_1,\lambda_2} \,\,
  d^{J}_{\sigma,\,\lambda_1-\lambda_2}(\theta)\,
  e^{i(\sigma-\lambda_1+\lambda_2)\phi}
  \quad\ \ \mbox{with}\quad\ \
  |\lambda_1-\lambda_2|\leq J\,,
\label{eq:x_m1m2_helicity_amplitude}
\end{eqnarray}
with the constraint $|\lambda_1-\lambda_2|\leq J$ in the JW
convention~\cite{Jacob:1959at,Haber:1994pe} (see
Figure~\ref{fig:kinematic_configuration_xm1m2_xrf} for the kinematic configuration).
Here, the helicity $\sigma$ of the spin-$J$ massive particle $X$ takes
one of $2J+1$ values between $-J$ and $J$. In contrast, each of the
helicities, $\lambda_{1,2}$, can take one of $2s+1$ values between $-s$ and
$s$ in the massive ($m\neq 0$) case but they take just two
values of $\pm s$ in the massless ($m=0$) case, as only the
maximal-magnitude helicity values identical to the spin in magnitude
are allowed physically for a massless particle of any spin.
The reduced helicity amplitudes ${\cal C}^J_{\lambda_1,\lambda_2}$
in Eq.~(\ref{eq:x_m1m2_helicity_amplitude}) do not depend on the $X$ helicity
$\sigma$ due to rotational invariance and the polar-angle dependence is
fully encoded
in the Wigner $d$ function $d^{J}_{\sigma,\lambda_1-\lambda_1}(\theta)$ given
in the convention of Rose~\cite{merose2011}.\s

If two particles, $M_1$ and $M_2$, are identical, the Bose or Fermi
symmetry in the integer or half-integer case leads to the IP relation
for the reduced helicity amplitudes:
\begin{eqnarray}
  {\cal C}^J_{\lambda_1,\lambda_2}
\, \, =\, \, (-1)^{J}\,\, {\cal C}_{\lambda_2,\lambda_1}
\ \ \ \ \mbox{with}\ \ \ \
  |\lambda_1-\lambda_2|\leq J\,,
\label{eq:x_m1m2_identical_condition}
\end{eqnarray}
due to the (anti)-symmetrization of the two identical
final-state particles.\footnote{Other discrete symmetries
like parity invariance put their corresponding constraints on the reduced
helicity amplitudes, although none of them are considered
in the present work.}\s

First, {\it in the massive ($m\neq 0$) case}, the number $n\,[J,s]$ of
independent reduced helicity amplitudes for specific $J$ and $s$ is
\begin{eqnarray}
  n\,[J,s]
\, =\,
  \left\{\begin{array}{ll}
         (2s+1)^2  & \ \ \mbox{for}\ \ J\geq 2s\,, \\[3mm]
         2s+1+(4s+1)J - J^2 & \ \  \mbox{for} \ \  J < 2s\,.
         \end{array}
  \right.
\label{eq:number_of_general_independent_terms}
\end{eqnarray}
For example, we have $n\,[J,0]=1$, $n\,[1,1]=7$ and $n\,[2,1]=9$.
On the other hand, the number of independent terms in the IP case
reduces to
\begin{eqnarray}
  n\,[J,s]_{\rm IP}
\, =\,
  \left\{\begin{array}{ll}
         \frac{1}{2}(2s+1)\left[1+(-1)^J\right]+s(2s+1)
           & \ \ \mbox{for} \ \ J\geq 2s \,, \\[4mm]
         \frac{1}{2}(2s+1)\left[1+(-1)^J\right]
         + \frac{1}{2}\left[(4s+1)J - J^2\right]
           & \ \ \mbox{for} \ \  J < 2s\,,
         \end{array}
  \right.
\label{eq:number_of_IP_independent_terms}
\end{eqnarray}
which depends crucially on whether the $X$ spin $J$ is even or
odd. For example, $n\,[J,0]_{\rm IP}=[1+(-1)^J]/2$,
$n\,[1,1]_{\rm IP}=2$ and $n\,[2,1]_{\rm IP}= 6$. Therefore,
any particle with an odd spin $J$ cannot decay into two identical
spinless particles.\s

In contrast, {\it in the massless ($m=0$) case}, the number $n\,[J,s]$
of independent reduced helicity amplitudes is reduced to
\begin{eqnarray}
 n\,[J,s]
\, =\,
  \left\{\begin{array}{ll}
         4  & \ \ \mbox{for}\ \ J\geq 2s \\[3mm]
         2 & \ \  \mbox{for} \ \  J < 2s
         \end{array}
  \right. \ \ \mbox{for} \ \ s>0
   \quad \mbox{and} \quad
   n\,[J,0]
\, =\, 1\,.
\label{eq:number_of_general_massless_independent_terms}
\end{eqnarray}
The number of independent terms in the case with $s=0$ is one,
irrespective of the $X$ spin $J$. On the other hand
the number of independent terms in the IP case is further reduced to
\begin{eqnarray}
 n\,[J,s]_{\rm IP}
\, =\,
  \left\{\begin{array}{ll}
         2+(-1)^J  & \ \ \mbox{for}\ \ J\geq 2s \\[3mm]
         1+(-1)^J & \ \  \mbox{for} \ \  J < 2s
         \end{array}
  \right. \ \ \mbox{for} \ \ s>0
   \quad \mbox{and} \quad
   n\,[J,0]
\, =\, \frac{1}{2}\,[ 1+(-1)^J]\,,
\label{eq:number_of_IP_massless_independent_terms}
\end{eqnarray}
due to the Bose or Fermi symmetry. One immediate consequence is that
any odd-$J$ particle cannot decay into two identical massless particles
of spin $s$ larger than $J/2$~\cite{Choi:2021ewa}. One well-known example
is that the decay of a spin-1 particle into two identical spin-1 massless
particles like two photons~\cite{Landau:1948kw,Yang:1950rg}.\s

\setcounter{equation}{0}

\section{Spin-$J$ and spin-$s$ wave tensors}
\label{sec:spin-j_spin-s_wave_tensors}

\begin{figure}[htb]
\begin{center}
\includegraphics[width=9.0cm, height=5.5cm]{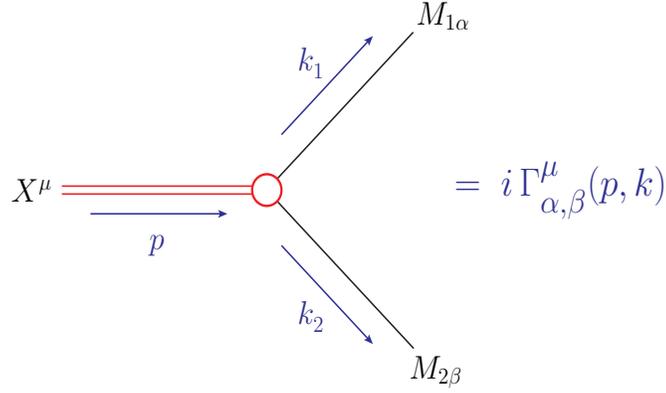}
\caption{\it Feynman rules $\Gamma^\mu_{\alpha,\beta}(p,k)$ for the
             general $XM_1M_2$ three-point vertex of a spin-$J$ particle
             $X$ of mass $m_X$ and two massive particles $M_1$ and $M_2$
             of equal mass $m$ and spin $s$.
             The indices, $\mu$, $\alpha$ and $\beta$, stand for the
             sequences of $\mu=\mu_1\cdots \mu_{_J}$,\,
             $\alpha_1\cdots \alpha_{n}$ and $\beta_1\cdots\beta_{n}$
             collectively with the non-negative integer $n=s$ or $n=s-1/2$
             in the integer or half-integer spin $s$ case.
             The symmetric and anti-symmetric momentum
             combinations, $p=k_1+k_2$ and $k=k_1-k_2$, are introduced
             for systematic classifications of the three-point
             vertex tensor.
}
\label{fig:xm1m2_vertex}
\end{center}
\end{figure}

Generically, the decay amplitude of one spin-$J$ particle $X$ of mass
$m_X$ into two particles, $M_1$ and $M_2$, of
equal mass $m$ and spin $s$, can be written in terms of the
three-point vertex tensor $\Gamma$ (see Figure~\ref{fig:xm1m2_vertex}
for its diagrammatic description)
\begin{eqnarray}
    {\cal M}^{X\to M_1M_2}_{\sigma; \lambda_1,\lambda_2}
&=& \bar{u}_1^{\alpha_1\cdots\alpha_{n}}(k_1,\lambda_1)\,\,
    \Gamma^{\mu_1\cdots\mu_J}_{\alpha_1\cdots\alpha_n,\beta_1\cdots\beta_n}(p,k)
     \,\, v_2^{\beta_1\cdots\beta_{n}}(k_2,\lambda_2)\,\,
    \epsilon_{\mu_1\cdots\mu_J}(p,\sigma)\,,
\label{eq:xm1m2_vertex}
\end{eqnarray}
with the non-negative integer $n=s$ or $n=s-1/2$ in the integer
or half-integer spin $s$ case, respectively. $p$ and $\sigma$ are the
momentum and helicity of the particle $X$,
and $k_{1,2}$ and  $\lambda_{1,2}$ are the momenta and helicities of two particles, $M_1$ and $M_2$, respectively. Here, $p=k_1+k_2$ and $k=k_1-k_2$
are symmetric and anti-symmetric under the interchange of two momenta,
$k_1$ and $k_2$. \s

\vskip 0.5cm

If the spin $s$ is an integer, then the bosonic wave tensors
$\bar{u}_1^{\alpha_1\cdots\alpha_s}(k_1,\lambda_1)$
and $v_2^{\beta_1\cdots\beta_s}(k_2,\lambda_2)$ for the particles
with non-zero mass $m$ are given by
\begin{eqnarray}
       \bar{u}_1^{\alpha_1\cdots\alpha_s}(k_1,\lambda_1)
\, &=& \, \epsilon_1^{*\alpha_1\cdots\alpha_s}(k_1,\lambda_1)\,, \\
       v_2^{\beta_1\cdots\beta_s}(k_2,\lambda_2)
\, &=& \, \epsilon_2^{*\beta_1\cdots\beta_s}(k_2,\lambda_2)\,,
\label{general_wave_tensors}
\end{eqnarray}
each of which is explicitly given by~\cite{Behrends:1957rup,Auvil:1966eao,
Caudrey:1968vih,Scadron:1968zz,Chung:1997jn,Huang:2003ym,
Miyamoto:2011aa}
\begin{eqnarray}
  \epsilon^{*\alpha_1\cdots\alpha_s}_1(k_1,\lambda_1)
& = &
   \sqrt{\frac{2^s\,(s+\lambda_1)!\, (s-\lambda_1)!}{(2s)!}}\,
   \sum^{1}_{\{\tau\}=-1}\,
   \delta_{\tau_1+\cdots+\tau_s,\,\lambda_1}\,
   \prod_{i=1}^s\,
   \frac{\epsilon_1^{*\alpha_i}(k_1,\tau_i)}{
          \sqrt{2}^{|\tau_i|}}\,,\\
\epsilon^{*\beta_1\cdots\beta_s}_2(k_2,\lambda_2)
& = &
   \sqrt{\frac{2^s\,(s+\lambda_2)!\, (s-\lambda_2)!}{(2s)!}}\,
   \sum^{1}_{\{\tau\}=-1}\,
   \delta_{\tau_1+\cdots+\tau_s,\,\lambda_2}\,
   \prod_{i=1}^s\,
   \frac{\epsilon_2^{*\beta_i}(k_2,\tau_i)}{
         \sqrt{2}^{|\tau_i|}}\,,
\label{eq:explicit_bosonic_wave_tensors}
\end{eqnarray}
with the convention $\{\tau\}=\tau_1,\cdots,\tau_s$ as a linear
combination of products of $s$ polarization vectors with appropriate Clebsch-Gordon coefficients.
We note that the bosonic wave tensors are {\it totally symmetric,
traceless and divergence-free}
\begin{eqnarray}
      \varepsilon_{\mu\nu\alpha_i\alpha_j}\,
      \epsilon_{a}^{\alpha_1\cdots\alpha_i\cdots\alpha_j\cdots\alpha_s}
       (k_{a},\lambda_{a})
&=& 0\,,
   \label{eq:totally_symmetric}\\[1mm]
      g_{\alpha_i\alpha_j}\,
      \epsilon_{a}^{\alpha_1\cdots\alpha_i\cdots\alpha_j\cdots\alpha_s}
      (k_{a},\lambda_{a})
&=& 0\,,
   \label{eq:traceless}\\[1mm]
      k_{\alpha_i}\,
      \epsilon_{a}^{\alpha_1\cdots\alpha_i\cdots\alpha_s}
      (k_{a},\lambda_{a})
&=& 0\,,
   \label{eq:divergence_free}
 \end{eqnarray}
with $a=1$ and $2$, and both of them satisfy the same on-shell wave
equation $(k^2_{1,2}-m^2)\,
\epsilon_{1,2}^{\alpha_1\cdots\alpha_s}(k_{1,2},\lambda_{1,2})=0$
for any helicity value of $\lambda_{1,2}$ taking an integer between
$-s$ and $s$. In contrast, if $m=0$, the wave tensors with two
allowed helicities $\pm s$ are given simply by a direct product of
$s$ spin-1 polarization vectors, each of which carries the same helicity
of $\pm 1$, as
\begin{eqnarray}
&& \bar{u}_1(k_1,\pm s)
   \,=\, \epsilon_1^{*\alpha_1\cdots\alpha_s}(k_1,\pm s)
   \,=\, \epsilon_1^{*\alpha_1}(k_1,\pm 1)\cdots
         \epsilon_1^{*\alpha_s}(k_1,\pm 1)\,,\\
&& v_2(k_2,\pm s)
   \,\,=\, \epsilon_2^{*\beta_1\cdots\beta_s}(k_2,\pm s)
   \,\,=\, \epsilon_2^{*\beta_1}(k_2,\pm 1)\cdots
         \epsilon_2^{*\beta_s}(k_2,\pm 1)\,,
\label{eq:explicit_massless_wave_tensors}
\end{eqnarray}
which are totally symmetric, traceless and divergence-free in the
four-vector
indices, $\alpha=\alpha_1\cdots\alpha_s$ and
$\beta=\beta_1\cdots\beta_s$, as well.\s

On the other hand, for a half-integer $s=n+1/2$ with a non-negative
integer $n$, the fermionic wave tensors for the particles with non-zero
mass are given by~\cite{Behrends:1957rup,Auvil:1966eao,
Caudrey:1968vih,Scadron:1968zz,Huang:2003ym,Miyamoto:2011aa}
\begin{eqnarray}
&& \mbox{ }\hskip -1.3cm
  \bar{u}^{\alpha_1\cdots\alpha_n}_1(k_1,\lambda_1)
= \sum_{\tau=\pm\frac{1}{2}}\,
   \sqrt{\frac{s+2\tau \lambda_1}{2s}}\,\,
   \epsilon^{*\alpha_1\cdots\alpha_n}_1
    (k_1,\lambda_1-\tau)\,
   \bar{u}(k_1,\tau)\,,\\
&& \mbox{ }\hskip -1.3cm
  v^{\alpha_1\cdots\alpha_n}_2(k_2,\lambda_2)
= \sum_{\tau=\pm \frac{1}{2}}\,
   \sqrt{\frac{s+2\tau\lambda_2}{2s}}\,\,
   \epsilon^{*\alpha_1\cdots\alpha_n}_2
    (k_2,\lambda_2-\tau)\,
   v_2(k_2,\tau)\,,
\label{eq:explicit_fermionic_wave_spinors}
\end{eqnarray}
where $\bar{u}_1(k_1,\mbox{\small $\pm\frac{1}{2}$})
=u^\dagger_1(k_1,\mbox{\small $\pm\frac{1}{2}$})\gamma^0$ with
the spin-1/2 particle spinor $u_1(k_1,\mbox{\small $\pm\frac{1}{2}$})$,
and $v_2(k_2,\mbox{\small $\pm\frac{1}{2}$})$ is the spin-1/2
anti-particle spinor. The spin-1/2 spinors
satisfy their own on-shell equations,
$(\not\!\!{k}_1-m)u_1(k_1,\mbox{\small $\pm\frac{1}{2}$})=0$
and $(\not\!\!{k}_2+m)v_2(k_2,\mbox{\small $\pm\frac{1}{2}$})=0$.
In contrast, if $m=0$, the wave tensors with two allowed helicities
$\pm s$ are given by a product of a $u$ or $v$ spinor and
$n$ spin-1 wave vectors with $n=s-1/2$ as
\begin{eqnarray}
&& \bar{u}_1^\alpha(k_1,\pm s)
 \, =\, \epsilon_1^{*\alpha_1}(k_1,\pm 1)\cdots
        \epsilon_1^{*\alpha_n}(k_1,\pm 1)\,
        \bar{u}_1(k_1, \pm\mbox{\small $\frac{1}{2}$})\,, \\
&& v_2^\beta(k_2,\pm s)
 \, =\, \epsilon_2^{*\beta_1}(k_2,\pm 1)\cdots
        \epsilon_2^{*\beta_n}(k_2,\pm 1)\,
        v_2(k_2, \pm\mbox{\small $\frac{1}{2}$})\,,
\label{eq:explicit_massless_wave_spinors}
\end{eqnarray}
which are totally symmetric, traceless and divergence-free
in the four-vector indices, $\alpha=\alpha_1\cdots\alpha_n$ and
$\beta=\beta_1\cdots\beta_n$, as well. \s

Similarily, the on-shell boson $X$ of an integer spin $J$, mass
$m_X$, momentum $p$ and helicity $\sigma$ is represented by
a totally-symmetric, traceless and divergence-free rank-$J$ bosonic wave tensor $\epsilon_{\mu_1\cdots\mu_{_J}}(p,\sigma)$~\cite{Behrends:1957rup,Auvil:1966eao,
Caudrey:1968vih,Scadron:1968zz,Chung:1997jn,Huang:2003ym,
Miyamoto:2011aa}. The explicit form of the bosonic wave tensor is
given by
\begin{eqnarray}
\epsilon_{\mu_1\cdots\mu_J}(p,\sigma)
& = &
   \sqrt{\frac{2^J\,(J+\sigma)!\, (J-\sigma)!}{(2J)!}}\,
   \sum^{1}_{\{\tau\}=-1}\,
   \delta_{\tau_1+\cdots+\tau_J,\,\sigma}\,
   \prod_{i=1}^J\,
   \frac{\epsilon_{\mu_i}(p,\tau_i)}{
         \sqrt{2}^{|\tau_i|}}\,,
\label{eq:explicit_x_wave_tensors}
\end{eqnarray}
with the convention $\{\tau\}=\tau_1,\cdots,\tau_s$, which satisfies
the on-shell equation of motion
$(p^2-m^2_X)\epsilon_{\mu_1\cdots\mu_J}(p,\sigma)=0$ for any helicity
value of $\sigma$ taking an integer value between $-J$ and $J$. \s

For calculating the reduced helicity amplitudes in the following,
we show the explicit expressions for the wave vectors and spinors
of the particle $X$ and two particles $M_{1,2}$ in the $X$RF
with the kinematic configuration as shown
in Figure~\ref{fig:kinematic_configuration_xm1m2_xrf}.
The JW convention of Refs.~\cite{Jacob:1959at,Haber:1994pe} is
adopted for the vectors and spinors. For the sake of notation,
we introduce three unit vectors
\begin{eqnarray}
   \hat{n}
&=& (\sin\theta\cos\phi,\, \sin\theta\sin\phi,\, \phantom{+}\cos\theta)\,, \\
   \hat{\theta}
&=& (\cos\theta\cos\phi,\, \cos\theta\sin\phi,\, -\sin\theta) \,, \\
   \hat{\phi}
&=& (-\sin\phi,\, \cos\phi,\, 0)\,,
\label{eq:three_orthonormal_unit_vectors}
\end{eqnarray}
expressed in terms of the polar and azimuthal angles, $\theta$ and $\phi$.
The three unit vectors are mutually orthonormal, i.e.
$\hat{n}\cdot\hat{\theta}=\hat{\theta}\cdot\hat{\phi}
=\hat{\phi}\cdot\hat{n}=0$ and $\hat{n}\cdot\hat{n}=\hat{\theta}\cdot
\hat{\theta}=\hat{\phi}\cdot\hat{\phi}=1$.
The four-momentum sum $p=k_1+k_2$ and the four-momentum difference
$k=k_1-k_2$ read
\begin{eqnarray}
p \, =\, m_X \hat{p}\, =\, m_X\, (1, \vec{0}) \quad \mbox{and} \quad
k\, =\, m_X\kappa \hat{k}\, =\,m_X\kappa\,\, (0, \hat{n})\,,
\label{eq:explicit_p_k_momenta_xrf}
\end{eqnarray}
with $\kappa=\sqrt{1-4m^2/m^2_X}$ the speed of the particles $M_{1,2}$
in the $X$RF. In the following, we use the normalized momenta
$\hat{p}=(1,\vec{0})$ and $\hat{k}=(0,\hat{n})$ for calculating all the
reduced helicity amplitudes in the $X$RF.\s

The spin-1 wave vectors for the particle $X$ with momentum $p$ and two
particles $M_{1,2}$
with momenta $k_{1,2}=(p\pm k)/2=m_X(1,\pm\kappa \hat{n})/2$ are given
in the JW convention by
\begin{eqnarray}
\epsilon(p,\pm 1) \! &=&\!
     \frac{1}{\sqrt{2}}\, (0,\, \mp 1,\,  -i,\, 0)\,, \ \ \ \ \ \ \
\epsilon(p,\, 0) \,\,\,\, =\,\,
     (0,\, 0,\, 0,\, 1)\,,
     \label{eq:xrf_x_wave_vectors} \\
\epsilon_1(k_1,\pm 1) \! &=&\!
     \frac{1}{\sqrt{2}}\, e^{\pm i \phi}\,
     (0,\, \mp \hat{\theta}-i \hat{\phi})\,, \ \ \
\epsilon_1(k_1, 0) \, =\,
     \frac{m_X}{2m}\,
     (\phantom{+}\kappa,\, \hat{n})\,,
     \label{eq:xrf_m1_wave_vectors} \\
\epsilon_2(k_2,\pm 1) \! &=&\!
     \frac{1}{\sqrt{2}}\, e^{\mp i \phi}\,
     (0,\, \pm \hat{\theta}-i \hat{\phi})\,, \ \ \
\epsilon_2(k_2, 0) \, =\,
     \frac{m_X}{2m}\,
     (-\kappa,\, \hat{n})\,,
     \label{eq:xrf_m2_0_wave_vector}
\end{eqnarray}
in the $X$RF, among which the transverse wave vectors satisfy the relation,
$\epsilon_2(k_2,\pm 1) = \epsilon_1(k_1, \mp 1)
=  -\epsilon^*_1(k_1,\pm 1) =  -\epsilon^*_2(k_2,\mp 1)$
in the JW convention. \s

The spin-1/2 $u_1$ and $v_2$ spinors of the particles $M_{1,2}$
are given in the JW convention by
\begin{eqnarray}
u_1(k_1,\pm\mbox{\small $\frac{1}{2}$})
   \,=\, \sqrt{\frac{m_X}{2}}\,
      \left(\begin{array}{c}
            \sqrt{1\mp\kappa}\, \chi_\pm(\hat{n}) \\[3mm]
            \sqrt{1\pm\kappa}\, \chi_\pm(\hat{n})
                   \end{array}\right)\quad \mbox{and}\quad
v_2(k_2,\pm\mbox{\small $\frac{1}{2}$})
  \, = \, \pm\sqrt{\frac{m_X}{2}}\,
      \left(\begin{array}{c}
            \sqrt{1\pm\kappa}\, \chi_\pm(\hat{n}) \\[3mm]
            -\sqrt{1\mp\kappa}\, \chi_\pm(\hat{n})
                   \end{array}\right)\,,
\label{eq:xrf_m_1_m_2_wave_spinors}
\end{eqnarray}
where the 2-component spinors $\chi_\pm(\hat{k})$ are written in terms of
the polar and azimuthal angles, $\theta$ and $\phi$, as
\begin{eqnarray}
\chi_+(\hat{k}) \,=\, \left(\begin{array}{l}
                       \cos\frac{\theta}{2}\, \\[2mm]
                       \sin\frac{\theta}{2}\, e^{i\phi}
                       \end{array}\right)\quad \mbox{and}\quad
\chi_-(\hat{k}) \,=\, \left(\begin{array}{c}
                       -\sin\frac{\theta}{2}\, e^{-i\phi} \\[2mm]
                       \!\!\!\!\!\!\cos\frac{\theta}{2} \,
                       \end{array}\right)\,,
\label{eq:2-component_spinor_wave_functions}
\end{eqnarray}
being mutually orthonormal, i.e. $\chi^\dagger_a(\hat{k})
\chi_b(\hat{k})=\delta_{a,b}$, with $a,b=\pm$, in the $X$RF.\s

\setcounter{equation}{0}

\section{Basic covariant three-point vertices}
\label{sec:basic_covariant_three_point_vertices}

In this section, we derive all the Lorentz-covariant operators
corresponding to the reduced helicity amplitudes for three values
of $J=0,1,2$ and a fixed value of $s=1$. Those covariant operators
constitute the backbone for weaving the covariant three-point vertices
for arbitrary $J$ and $s$.\s

\subsection{Bosonic vertex operators}
\label{subsec:bosonic_vertex_operators}

First, we consider the decay of a spin-0 particle $X$ into
two spin-1 bosons, $M_1$ and $M_2$. The number of independent
terms involving the $0\to 1+1$ decay is $n\,[0,1]=3$, accounting
for the three reduced helicity amplitudes, ${\cal C}^{\,0}_{\pm 1,\pm 1}$
and ${\cal C}^{\,0}_{0,0}$, in the $X$RF. After a little manipulation,
we can find the three covariant three-point vertex operators defined as
\begin{eqnarray}
&& S^\pm_{\alpha\beta}
  \, =\, \frac{1}{2}\, \left[g_{\bot\alpha\beta} \pm i \langle\alpha\beta\hat{p}\hat{k}\rangle\right]
  \ \ \leftrightarrow\qquad {\cal C}^{\,0}_{\pm 1,\pm 1} = \, 1\,,
  \label{eq:bosonic_scalar_composite_operators_pm}\\
&& S^0_{\alpha\beta}
  \, =\, \frac{4m^2}{m^2_X}\,\hat{p}_\alpha\hat{p}_\beta
  \qquad\quad\ \ \ \ \,\, \,\,\, \leftrightarrow\qquad
  {\cal C}^{\,0}_{0,0}\,\,\ \ =\, -\kappa^2\,,
  \label{eq:bosonic_scalar_composite_operators_0}
\end{eqnarray}
with the orthogonal tensor $g_{\bot\alpha\beta} =g_{\alpha\beta}
-\hat{p}_\alpha\hat{p}_\beta +\hat{k}_\alpha\hat{k}_\beta$ and
$\langle \alpha\beta\hat{p}\hat{k}\rangle
=\varepsilon_{\alpha\beta\rho\sigma}
\hat{p}^\rho\hat{k}^\sigma$ in terms of the totally antisymmetric
Levi-Civita tensor with the sign convention $\varepsilon_{0123}=+1$.
Each of the three covariant
three-point vertices generates solely its corresponding reduced helicity
amplitude, as shown in Eqs.~(\ref{eq:bosonic_scalar_composite_operators_pm})
and (\ref{eq:bosonic_scalar_composite_operators_0}).\s

Second, there are in general $n\,[1,1]= 7$ independent terms for the
$1\to 1 +1$ decay mode, among which
three generate the same helicity combinations as in the case with
$J=0$. The corresponding covariant three-point vertices are simply
$\hat{k}_\mu\, S^\pm_{\alpha\beta}$ and $\hat{k}_\mu\, S^{0}_{\alpha\beta}$
generating their corresponding reduced helicity amplitudes,
${\cal C}^1_{\pm 1,\pm 1}=-1$ and ${\cal C}^1_{0,0}=\kappa^2$, which
are identical to $-{\cal C}^{\,0}_{\pm 1, \pm 1}$ and
$-{\cal C}^{\,0}_{0,0}$, respectively. The remaining four
covariant vertices and their corresponding reduced helicity amplitudes
are given by
\begin{eqnarray}
&&    V^\pm_{1\alpha\beta;\mu}
  \,=\, \frac{m}{m_X} \hat{p}_\beta
     \left[g_{\bot\alpha\mu}
           \pm i \langle \alpha\mu\hat{p}\hat{k}\rangle
     \right]
     \qquad\quad \leftrightarrow \qquad\quad
     {\cal C}^{1}_{\pm 1,0} = \phantom{+}\kappa\,,
  \label{eq:bosonic_vector_composite_operators_1}\\
&&    V^\pm_{2\alpha\beta;\mu}
  \,=\, \frac{m}{m_X} \hat{p}_\alpha
     \left[g_{\bot\beta\mu}
           \mp i \langle \beta\mu\hat{p}\hat{k}\rangle
     \right]
     \qquad\quad \leftrightarrow \qquad\quad
     {\cal C}^{1}_{0, \pm 1} = - \kappa\,,
  \label{eq:bosonic_vector_composite_operators_2}
\end{eqnarray}
with the orthogonal tensors, $g_{\bot\alpha\mu}
=g_{\alpha\mu}-\hat{p}_\alpha\hat{p}_\mu + \hat{k}_\alpha\hat{k}_\mu$ and
$g_{\bot\beta\mu}= g_{\beta\mu}-\hat{p}_\beta\hat{p}_\mu
  + \hat{k}_\beta\hat{k}_\mu$.\footnote{Making a suitable use of
Schouten identities, we can check that the set of seven three-point
vertices listed above are equivalent to that of seven $VW^+W^-$
three-point vertex terms listed in Ref.~\cite{Hagiwara:1986vm}.}\s

Third, the decay of a spin-2 particle $X$ into two spin-1 particles,
$M_1$ and $M_2$, is in general described by $n\,[2,1]=9$ independent
terms. Seven of them can be constructed simply by multiplying the
seven vertices participating in the spin-1 case by $\hat{k}$, i.e.,
$\hat{k}_{\mu_2}\hat{k}_{\mu_1} S^{\pm, 0}_{\alpha\beta}$,
$\hat{k}_{\mu_2} V^{\pm}_{1\alpha\beta;\mu_1}$ and
$\hat{k}_{\mu_2} V^{\pm}_{2\alpha\beta;\mu_1}$, generating the
reduced helicity amplitudes, ${\cal C}^{2}_{\pm 1,\pm 1} = 1$,
${\cal C}^{2}_{0,0}=-\kappa^2$, and ${\cal C}^{2}_{\pm 1, 0}
=-{\cal C}^{2}_{0,\pm 1}=-\kappa$, identical to
$-{\cal C}^{1}_{\pm 1,\pm 1}$,
$-{\cal C}^{1}_{0,0}$, $-{\cal C}^{1}_{\pm 1, 0}$, and
$-{\cal C}^{1}_{0,\pm 1}$, respectively.
The two remaining covariant vertices and their corresponding reduced
helicity amplitudes are
\begin{eqnarray}
     T^{\pm}_{\alpha\beta;\mu_1\mu_2}
\,=\,\frac{1}{4}\,
     \left[g_{\bot\alpha\mu_1}
          \pm i \langle \alpha\mu_1\hat{p}\hat{k}\rangle\right]
     \left[g_{\bot\beta\mu_2}
          \mp i \langle \beta\mu_2\hat{p}\hat{k}\rangle\right]
     \qquad \leftrightarrow \qquad
     {\cal C}^2_{\pm 1, \mp 1} = 1\,.
\label{eq:bosonic_tensor_composite_operators_pm}
\end{eqnarray}
Note that the number of independent terms does not increase any
more for $J$ larger than 2, i.e. $n\,[J,1]=9$ for $J\geq 2$.
Generally, the covariant three-point vertex
$\Gamma_{\alpha\beta;\mu_1\mu_2\cdots\mu_J}
=\Gamma_{\alpha\beta;\mu_1\mu_2}\hat{k}_{\mu_3}\cdots\hat{k}_{\mu_J}$
for $J\geq 2$. \s

Scrutinizing the structure of all the scalar, vector and tensor composite
vertex operators listed in
Eqs.~(\ref{eq:bosonic_scalar_composite_operators_pm}),
(\ref{eq:bosonic_scalar_composite_operators_0}),  (\ref{eq:bosonic_vector_composite_operators_1})  (\ref{eq:bosonic_vector_composite_operators_2}) and
(\ref{eq:bosonic_tensor_composite_operators_pm})
carefully, we realize that any non-trivial helicity
shifts in the $X$RF are generated essentially by two basic vector
operators, $U^+_{1,2}$, and their complex conjugates, $U^-_{1,2}$,
responsible for the positive and negative one-step transition of
the $M_1$ and $M_2$ helicity states as
\begin{eqnarray}
&& U^\pm_{1\alpha\mu}
 \, =\,\frac{1}{2}\left[g_{\bot\alpha\mu}
                       \pm i \langle \alpha\mu\hat{p}\hat{k}\rangle
                  \right]
   \qquad \Leftrightarrow \qquad
   [\lambda_1,\lambda_2]\ \ \rightarrow \ \ [\lambda_1\pm 1,\lambda_2]\,,\\
&& U^\pm_{2\beta\mu}
 \, =\,\frac{1}{2}\left[g_{\bot\beta\mu}
                       \mp i \langle \beta\mu\hat{p}\hat{k}\rangle
                 \right]
   \qquad \Leftrightarrow \qquad
   [\lambda_1,\lambda_2]\ \ \rightarrow \ \ [\lambda_1,\lambda_2\pm 1]\,,
 \label{eq:basic_vector_operators}
\end{eqnarray}
respectively.
In the operator form, the scalar and tensor composite three-point
vertex operators in Eqs.~(\ref{eq:bosonic_scalar_composite_operators_pm})
and (\ref{eq:bosonic_tensor_composite_operators_pm}) can be expressed
in an inner product and an outer product
of the operators, $U^\pm_1$ and $U^\pm_2$, as
\begin{eqnarray}
&&     S^\pm_{\alpha\beta}
  \,=\, g^{\mu_1\mu_2} U^{\pm}_{1\alpha\mu_1} U^{\pm}_{2\beta\mu_2}
  \,\equiv\, \left[U^\pm_1\cdot U^\pm_2\right]_{\alpha\beta}
  \qquad\qquad \Leftrightarrow \qquad
  [\lambda_1,\lambda_2] \ \ \rightarrow \ \
  [\lambda_1\pm 1, \lambda_2\pm 1]\,, \\
&&     T^\pm_{\alpha\beta;\mu_1\mu_2}
  \,=\, U^{\pm}_{1\alpha\mu_1} U^{\mp}_{2\beta\mu_2}
\,\,\equiv\,\, \left[U^\pm_1\star U^\mp_2\right]_{\alpha\beta;\mu_1\mu_2}
  \qquad \Leftrightarrow \qquad
  [\lambda_1,\lambda_2] \ \ \rightarrow \ \
  [\lambda_1\pm 1, \lambda_2\mp 1]\,,
  \label{eq:inner_outer_products}
\end{eqnarray}
respectively. Furthermore, in addition to the normalized momentum
$\hat{k}_\mu$, the momenta, $\hat{p}_{\alpha, \beta}$,  can be used
for matching the numbers of $\mu$-, $\alpha$- and $\beta$-type indices
for given $J$ and $s$, while keeping the helicity values intact,
and for defining the scalar and vector three-point vertices as
\begin{eqnarray}
     S^{0}_{\alpha\beta}
\,=\, \frac{2m}{m_X}\hat{p}_\alpha
      \frac{2m}{m_X}\hat{p}_\beta\,, \quad
     V^\pm_{1\alpha\beta;\mu}
  \,= \, \frac{2m}{m_X}\hat{p}_\beta\, U^\pm_{1\alpha\mu}
  \quad \mbox{and}\quad
 V^\pm_{2\alpha\beta;\mu}
  \,= \, \frac{2m}{m_X}\hat{p}_\alpha\, U^\pm_{2\beta\mu}\,.
  \label{eq:mass_dependent_composite_operators}
\end{eqnarray}
Clearly, these five composite operators in
Eq.~(\ref{eq:mass_dependent_composite_operators}) do not contribute
to the decay dynamics in the massless case with $m=0$, consistent
with the point that all the helicity-zero longitudinal modes are
absent for any massless states. \s

\begin{figure}[H]
\begin{center}
\includegraphics[width=6.0cm, height=5.5cm]{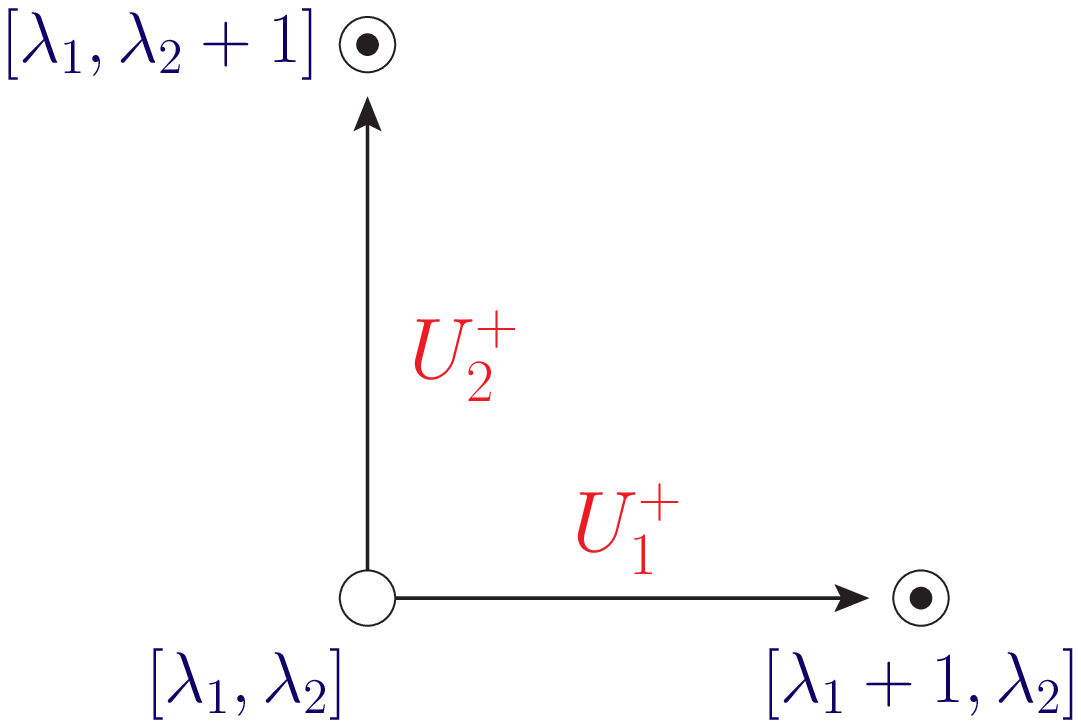}
\hskip 1.5cm
\includegraphics[width=6.5cm, height=6.5cm]{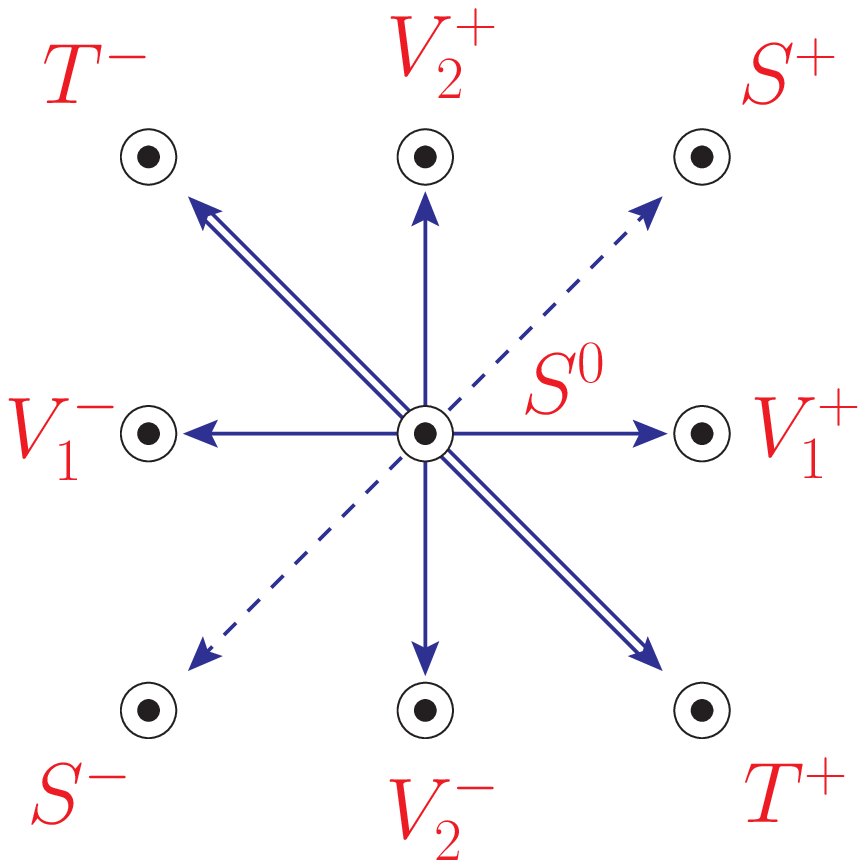}
\caption{\it (Left) A diagrammatic description of the basic operators,
             $U^+_1$ and $U^+_2$ moving the integer-helicity point
             $[\lambda_1,\lambda_2]$ to the integer-helicity point
             $[\lambda_1+1,\lambda_2]$ horizontally
             and to the integer-helicity point $[\lambda_1,\lambda_2+1]$
             vertically in the helicity-lattice space, respectively.
             Although not presented, the transitions moving down the
             integer-helicity point by one-step horizontally and
             vertically are dictated by the complex-conjugate operators,
             $U^-_1=(U^+_1)^*$ and $U^-_2=(U^+_2)^*$, respectively.
             (Right) A graphical description of the nine different transitions
             by the three scalar composite operators $S^{\pm}$ and $S^0$,
             four vector composite operators $V^{\pm}_1$ and $V^\pm_2$ and
             two tensor composite operators $T^\pm$.
}
\label{fig:basic_composite_transition_operators}
\end{center}
\end{figure}

In order to clarify the characteristics of the basic and composite
operators, let us introduce an integer-helicity lattice space consisting
of $(2s+1)\times (2s+1)$ in order for each point
$[\lambda_1,\lambda_2]$ to stand for its corresponding
reduced helicity amplitude ${\cal C}^J_{\lambda_1,\lambda_2}$
existing only when $|\lambda_1-\lambda_2|\leq J$ and
$|\lambda_{1,2}|\leq s$. As shown in the left panel of
Figure~\ref{fig:basic_composite_transition_operators},
the one-step increasing horizontal and vertical transitions
are dictated by the basic operators, $U^+_1$ and $U^+_2$,
from the point $[\lambda_1,\lambda_2]$ to the point
$[\lambda_1+1,\lambda_2]$ and the point $[\lambda_1,\lambda_2+1]$
in the helicity-lattice space, respectively.
By combining the normalized momenta $\hat{p}$ and $\hat{k}$ and
the basic transition operators properly, we can construct nine
different transitions consisting of three scalar composite operators
$S^{\pm,0}$, four vector composite operators $V^{\pm}_1$ and
$V^\pm_2$ and two tensor composite operators $T^\pm$, as shown
in the right panel of Figure~\ref{fig:basic_composite_transition_operators}.
Properly combining the nine composite operators enables us to reach
every integer-helicity lattice point. {\it To summarize, for any
given $J$ and $s$, we can weave the covariant three-point vertex
corresponding to every integer-helicity combination of
$[\lambda_1,\lambda_2]$ efficiently and systematically.}
The explicit form of every covariant
three-point vertex constructed by weaving the covariant composite operators
is to be presented in
Section~\ref{sec:weaving_covariant_three_point_vertices}. \s

\vskip 0.5cm

\subsection{Fermionic vertex operators}
\label{subsec:fermionic_vertex_operators}

First, we consider the decay of a spin-0 particle $X$ into
two spin-1/2 fermions, $M_1$ and $M_2$. The number of independent
terms involving the $0\to 1/2+1/2$ two-body decay is
$n\,[0,1/2]=2$, accounting for the two reduced helicity amplitudes,
${\cal C}^{\,0}_{\pm 1/2,\pm 1/2}$. After a little manipulation,
we can find the following two covariant three-point operators,
\begin{eqnarray}
P^\pm \, =\, \frac{1}{2m_X}\,(1\mp\kappa\gamma_5)
 \qquad \leftrightarrow\qquad
{\cal C}^{\,0}_{\pm 1/2, \pm 1/2} = \, \kappa\,,
\label{eq:fermion_scalar_three_point_operators}
\end{eqnarray}
with the $M_{1,2}$ speed $\kappa=\sqrt{1-4m^2/m^2_X}$ in the $X$RF.\s

Second, there are in general $n\,[1,1/2]= 4$ independent terms for the
$1\to  1/2 + 1/2$ decay
mode, among which two terms take the same helicity combinations as
in the $J=0$ case. The corresponding covariant three-point vertices
are simply $\hat{k}_\mu\, P^\pm$ generating their corresponding reduced
helicity amplitudes, ${\cal C}^1_{\pm 1/2,\pm 1/2}
=-\kappa $, identical to $-{\cal C}^{\,0}_{\pm 1/2,\pm 1/2}$.
The remaining two covariant vertices and their corresponding reduced
helicity amplitudes are given by
\begin{eqnarray}
&&    W^\pm_{\mu}
  \,=\, \frac{1}{2\sqrt{2}m_X}
   (\pm \kappa \gamma_{\bot\mu}  + \gamma_\mu \gamma_5)
     \qquad \leftrightarrow \qquad
     {\cal C}^{1}_{\pm 1/2,\mp 1/2} = \, \pm \kappa\,,
\label{eq:fermion_vector_three_point_operators}
\end{eqnarray}
with the orthogonal Dirac gamma matrix $\gamma_{\bot\mu}
=\gamma_\mu+ \frac{2m}{m_X\kappa}\,\hat{k}_\mu$.\s

\vskip 0.5cm

\begin{figure}[H]
\begin{center}
\includegraphics[width=6.cm, height=5.5cm]{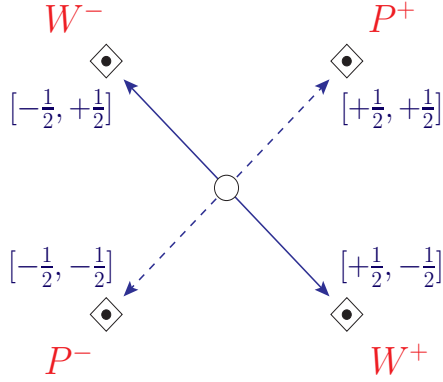}
\caption{\it A diagrammatic description of the half-step transitions
            by the four fermionic operators, $P^\pm$ and $W^\pm$ from
            the original point $[0,0]$ to the four half-integer helicity
            points $[\mbox{\small $\pm\frac{1}{2}$},\mbox{\small
            $\pm\frac{1}{2}$}]$ and $[\mbox{\small $\pm\frac{1}{2}$},
            \mbox{\small $\mp\frac{1}{2}$}]$, respectively.
}
\label{fig:fermionic_transition_operators}
\end{center}
\end{figure}

Properly combining the fermionic transition operators in
Eqs.~(\ref{eq:fermion_scalar_three_point_operators}) and
(\ref{eq:fermion_vector_three_point_operators}) and the
bosonic transition operators in
Eqs.~(\ref{eq:bosonic_scalar_composite_operators_pm}),
(\ref{eq:bosonic_scalar_composite_operators_0}),  (\ref{eq:bosonic_vector_composite_operators_1})  (\ref{eq:bosonic_vector_composite_operators_2}) and
(\ref{eq:bosonic_tensor_composite_operators_pm}) enables us to reach every half-integer-helicity
lattice point. {\it To summarize, for any given integer $J$ and
half-integer $s$, we can weave the covariant three-point vertex  corresponding to every half-integer-helicity combination of $[\lambda_1,\lambda_2]$
efficiently and systematically.} The explicit form of the covariant fermionic
three-point vertex constructed by weaving the fermionic as well as
bosonic operators is to be presented
in Section~\ref{sec:weaving_covariant_three_point_vertices}.\s

\setcounter{equation}{0}

\section{Weaving the covariant three-point vertices}
\label{sec:weaving_covariant_three_point_vertices}

Utilizing the nine composite bosonic operators and four basic
fermionic operators worked out in Section~\ref{sec:basic_covariant_three_point_vertices}, we weave
the covariant three-point vertices explicitly. For that purpose, it is
crucial to take into account that bosonic wave tensors are
totally symmetric, traceless and divergence-free and the
fermionic spinors satisfy
\begin{eqnarray}
    \gamma_{\alpha_i}
    u_1^{\alpha_1\cdots\alpha_i\cdots\alpha_n} (k_1,\lambda_1)
&=& \gamma_{\alpha_i} v_2^{\alpha_1\cdots\alpha_i\cdots\alpha_n}
    (k_2,\lambda_2)
\, =\,\, 0\,,
\label{eq:gamma_spinor_equation}
\end{eqnarray}
with the nonnegative integer $n=s-1/2$, so that every fermionic
vertex  involving $\gamma_{\alpha_i}$ or
$\gamma_{\beta_j}$ with $i,j=1,\cdots,n$  can be effectively excluded.
The $\mu$, $\alpha$ and $\beta$ four-vector indices in any
covariant three-point vertex can be  shuffled freely due to the totally
symmetric properties of the wave tensors, and any term including
$p_{\mu_i}$ for $i=1,\cdots J$ can be excluded effectively due to the
divergence-free condition. Moreover, the same condition allows us to
replace $k_{2\alpha_j}$ and $k_{1\beta_j}$ effectively
by $-p_{\alpha_j}$ and $p_{\beta_j}$ for $j=1,\cdots n$.\s

As many indices of different types are involved in expressing a
covariant three-point vertex especially for high-spin particles,
we introduce the following compact square-bracket notations
\begin{eqnarray}
&& [\,\hat{k}\,]^n
   \quad\ \ \rightarrow \quad
   (\hat{k}^n)_{\mu_1\cdots\mu_n}
   = \hat{k}_{\mu_1}\cdots\hat{k}_{\mu_n}\,,\\
&& [\,\hat{p}\,]^n
   \quad\ \ \rightarrow \quad
   (\hat{p}^n)_{\alpha_1\cdots\alpha_n}
   = \hat{p}_{\alpha_1}\cdots\hat{p}_{\alpha_n}
   \quad \mbox{and} \quad
   (\hat{p}^n)_{\beta_1\cdots\beta_n}
   = \hat{p}_{\beta_1}\cdots\hat{p}_{\beta_n}\,,\\
&& [S^\pm]^n
   \quad\, \rightarrow  \quad
   (S^\pm)^n_{\alpha_1\cdots\alpha_n\beta_1\cdots\beta_n}
   = S^\pm_{\alpha_1\beta_1}\cdots
     S^\pm_{\alpha_n\beta_n}\,, \\
&& [V_a^\pm]^n
  \quad \rightarrow \quad
   (V_a^\pm)^n_{\alpha_1\cdots\alpha_n\beta_1\cdots\beta_n;
                  \mu_1\cdots\mu_n}
   = V^\pm_{a\alpha_1\beta_1;\mu_1}\cdots
     V^\pm_{a\alpha_n\beta_n;\mu_n}\ \ \mbox{with}\ \  a=1,2\,,\\
&& [T^\pm]^n
  \quad \rightarrow \quad
   (T^\pm)^n_{\alpha_1\cdots\alpha_n\beta_1\cdots\beta_n;
                \mu_1\mu_2\cdots\mu_{2n-1}\mu_{2n}}
   = T^\pm_{\alpha_1\beta_1;\mu_1\mu_2}\cdots
     T^\pm_{\alpha_n\beta_n;\mu_{2n-1}\mu_{2n}}\,,
\end{eqnarray}
for a non-negative integer $n$. Obviously, the zeroth power ($n=0$) of
any operator or normalized four momentum is set to be 1.
We emphasize once more that any
permutation of the $\alpha$, $\beta$ and $\mu$ four-vector indices
can be regarded to be equivalent as eventually the vertex operators
are to be coupled with the $X$ and $M_{1,2}$ wave tensors totally
symmetric in the four-vector indices. \s

\subsection{Bosonic three-point vertices}
\label{subsec:bosonic_three_point_vertices}

In the integer $s$ case, any helicity lattice point where both $\lambda_1$
and $\lambda_2$ are even or odd can be reached through a sequence of
diagonal transitions by the scalar composite operators $S^\pm$ and $S^0$
and the tensor composite operators $T^\pm$. In fact, a little algebraic
manipulation leads to the following expression for the covariant
three-point vertex
\begin{eqnarray}
 [\,{\cal H}^{J,s}_{ii [\lambda_1,\lambda_2]}\,]
\, = \, [\,\hat{k}\,|^{J-2|\lambda_-|}\,
        [S^0]^{s-|\lambda_+|-|\lambda_-|}\,
        [S^{\hat{\lambda}_+}]^{|\lambda_+|}\,
        [T^{\hat{\lambda}_-}|^{|\lambda_-|}\,,
\label{eq:ii_helicity_specific_vertices}
\end{eqnarray}
in an operator form with two helicity combinations,
$\lambda_+=(\lambda_1+\lambda_2)/2$ and
$\lambda_-=(\lambda_1-\lambda_2)/2$, and their signs,
$\hat{\lambda}_+ ={\rm sign}(\lambda_+)$ and
$\hat{\lambda}_-={\rm sign}(\lambda_-)$.
We note that both $\lambda_+$ and $\lambda_-$ take integer values.
The subscript $ii$ implies that both $\lambda_+$ and $\lambda_-$ take
integer values. \s

On the other hand, any helicity lattice point where $\lambda_1$
and $\lambda_2$ are even and odd and vice versa can be reached
through a sequence of diagonal transitions by the scalar composite
operators $S^\pm$ and $S^0$ and the tensor composite operators
$T^\pm$ followed by a proper vertical or horizontal transition among
the vector composite operators $V^\pm_1$ and $V^\pm_2$. Explicitly,
we have the following expression with half-integer $\lambda_+$
and $\lambda_-$ for the three-point vertex
\begin{eqnarray}
 [\,{\cal H}^{J,s}_{hh [\lambda_1,\lambda_2]}\,]
\, = \, [\,\hat{k}\,|^{J-2|\lambda_-|}\,
        [S^0]^{s-|\lambda_+|-|\lambda_-|}\,
        [\,\delta_{\hat{\lambda}_+,\hat{\lambda}_-}\,
              V_1^{\hat{\lambda}_+}
              +\delta_{\hat{\lambda}_+,-\hat{\lambda}_-}
              V_2^{\hat{\lambda}_+}\,]\,
        [S^{\hat{\lambda}_+}]^{|\lambda_+|-1/2}\,
        [T^{\hat{\lambda}_-}|^{|\lambda_-|-1/2}\,,
\label{eq:hh_helicity_specific_vertices}
\end{eqnarray}
with two non-negative integer values of $|\lambda_\pm|-1/2$ in this case.
The subscript $hh$ implies that both $\lambda_+$ and $\lambda_-$ take
half-integer values.\s

\subsection{Fermionic three-point vertices}
\label{subsec:fermionic_three_point_vertices}

In the half-integer $s$ case with two fermions $M_1$ and $M_2$, the
helicity combinations can be categorized into two classes. One is
when the helicity sum $[\lambda_1+\lambda_2]$ and the helicity
difference $[\lambda_1-\lambda_2]$ are odd and even with a
half-integer $\lambda_+$ and an integer $\lambda_-$, to be
denoted by the notation $hi$. The other is when the helicity sum
and difference are even and odd with an integer $\lambda_+$ and
a half-integer $\lambda_-$, to be denoted by the notation $ih$.
Taking a proper half-integer helicity transition as described in
Figure~\ref{fig:fermionic_transition_operators} followed by
a sequence of integer helicity transitions, we can obtain
the following expression of the fermionic three-point vertex
with the helicities $\lambda_1$ and $\lambda_2$ as
\begin{eqnarray}
    [\, {\cal H}^{J,s}_{hi [\lambda_1,\lambda_2]}\,]
\,=\,   [\,\hat{k}\,]^{J-2|\lambda_-|}\,
        [S^0]^{s-|\lambda_+|-|\lambda_-|}\,
        [P^{\hat{\lambda}_+}]\,
        [S^{\hat{\lambda}_+}]^{|\lambda_+|-1/2}\,
        [T^{\hat{\lambda}_-}]^{|\lambda_-|}\,,
\label{eq:hi_helicity_specific_vertices}
\end{eqnarray}
involving a fermionic scalar operator $P^\pm$
for half-integer $\lambda_+$ and integer $\lambda_-$ and
\begin{eqnarray}
    [\, {\cal H}^{J,s}_{ih [\lambda_1,\lambda_2]}\,]
\,=\,   [\,\hat{k}\,]^{J-2|\lambda_-|}\,
        [S^0]^{s-|\lambda_+|-|\lambda_-|}\,
        [W^{\hat{\lambda}_-}]\,
        [S^{\hat{\lambda}_+}]^{|\lambda_+|}\,
        [T^{\hat{\lambda}_-}]^{|\lambda_-|-1/2}\,,
\label{eq:ih_helicity_specific_vertices}
\end{eqnarray}
involving a fermionic vector operator $W^\pm$
for integer $\lambda_+$ and half-integer $\lambda_-$.
The subscript $hi$ ($ih$) implies that $\lambda_+$ takes
a half-integer (an integer) value and $\lambda_-$ an integer
(a half-integer) value.\s

{\it To conclude}, in both the integer and half-integer spin $s$ cases,
the general form of any covariant three-point vertex
$\Gamma_{\alpha\beta;\mu}$ for given $J$ and $s$ is a linear combination
of all the allowed helicity-specific three-point vertices.
The succinct operator form of the covariant three-point vertex is
given by
\begin{eqnarray}
[\,\Gamma\,] \,=\, \left\{\begin{array}{ll}
             \sum^{s}_{\lambda_{1,2}=-s} \,
             \left[\, c^{J,s}_{ii [\lambda_1,\lambda_2]}\,
             [\, {\cal H}^{J,s}_{ii [\lambda_1,\lambda_2]}\,]
              + c^{J,s}_{hh [\lambda_1,\lambda_2]}\,
             [\, {\cal H}^{J,s}_{hh [\lambda_1,\lambda_2]}\,]\,
             \right] & \ \ \mbox{for an integer $s$}\,, \\[5mm]
             \sum^{s}_{\lambda_{1,2}=-s} \,
             \left[\, c^{J,s}_{hi [\lambda_1,\lambda_2]}\,
             [\, {\cal H}^{J,s}_{hi [\lambda_1,\lambda_2]}\,]
              + c^{J,s}_{ih [\lambda_1,\lambda_2]}\,
             [\, {\cal H}^{J,s}_{ih [\lambda_1,\lambda_2]}\,]\,
             \right] & \ \ \mbox{for a half-integer $s$}\,,
             \end{array}
             \right.
\label{eq:general_covariant_three_point_vertices}
\end{eqnarray}
with the constraint $|\lambda_1-\lambda_2|\leq J$, i.e.,
$J-2|\lambda_-|\geq 0$, where the helicity-specific coefficients,
$c^{J,s}_{x [\lambda_1,\lambda_2]}$ with $x=ii, hh, hi, hi$
depend only on the $X$ and $M_{1,2}$ masses.
We claim that the expression (\ref{eq:general_covariant_three_point_vertices})
along with the helicity-specific vertices in
Eqs.~(\ref{eq:ii_helicity_specific_vertices}),
(\ref{eq:hh_helicity_specific_vertices}),
(\ref{eq:hi_helicity_specific_vertices}) and
(\ref{eq:ih_helicity_specific_vertices}) is the key result of
the present work. Although it is originally deduced from the comparison
with the helicity amplitudes in the $X$RF, {\it the form is valid in
every reference frame because of its Lorentz-covariant form.} \s

\subsection{Massless case}
\label{subsec:massless_case}

In the case with massless $M_{1,2}$ of spin $s$, the physically-allowed
helicity values are $\pm s$. Furthermore, the five scalar and vector
operators $S^0$ and $V^\pm_{1,2}$ are vanishing as
they are proportional to $m^2$ and $m$, respectively. Therefore,
the helicity-specific vertices could survive only when
$\lambda_1=\lambda_2=\pm s$ or $\lambda_1=-\lambda_2=\pm s$. These
combinations satisfy $s-|\lambda_+|+|\lambda_-|=0$. In addition,
in the opposite helicity case, the $X$ spin $J$ cannot be smaller than
$2s$. Consequently, {\it in the $m=0$ case, there could exist at most
four independent terms in both the bosonic and fermionic cases}, as
counted in Eq.~(\ref{eq:number_of_general_massless_independent_terms}).\s

In the bosonic case with an integer $s$, the bosonic covariant three-point
vertex is given in an operator form by
\begin{eqnarray}
    [\, \Gamma\,]
&=&  c^{J,s}_{ii [+s,+s]}\, [\,k\,]^J\, [S^+]^s
   + c^{J,s}_{ii [-s,-s]}\, [\,k\,]^J\, [S^-]^s\nonumber\\
&+& \theta(J-2s)\,
    \left\{\, c^{J,s}_{ii [+s,-s]}\, [\,k\,]^{J-2s}\,[T^+]^s
            + c^{J,s}_{ii [-s,+s]}\, [\,k\,]^{J-2s}\, [T^-]^s\, \right\}\,,
\label{eq:massless_bosonic_three_point_vertex}
\end{eqnarray}
where the last two terms survive only when $J\geq 2s$, as denoted
by the step function $\theta(J-2s)=1$ for $J\geq 2s$ and $0$ for $J< 2s$.
On the other hand, in the
fermionic case with a half-integer $s$, the fermionic covariant
three-point vertex is given in an operator form as
\begin{eqnarray}
    [\,\Gamma\,]
&=&  c^{J,s}_{hi [+s,+s]}\, [\,k\,]^J\, [P^+]\,[S^+]^{s-1/2}
   + c^{J,s}_{hi [-s,-s]}\, [\,k\,]^J\, [P^-]\,[S^-]^{s-1/2} \nonumber\\
&+& \theta(J-2s)\,
    \left\{\, c^{J,s}_{ih [+s,-s]}\, [\,k\,]^{J-2s}\,
             [W^+]\, [T^+]^{s-1/2}
            + c^{J,s}_{ih [-s,+s]}\, [\,k\,]^{J-2s}\,
             [W^-]\, [T^-]^{s-1/2}\, \right\}\,.
\label{eq:massless_fermionic_three_point_vertex}
\end{eqnarray}
The results in Eqs.~(\ref{eq:massless_bosonic_three_point_vertex})
and (\ref{eq:massless_fermionic_three_point_vertex})
are consistent with those derived in Ref.~\cite{Choi:2021ewa}.\s

\subsection{Identical particle relation: Bose or Fermi symmetry}
\label{subsec:identical_particle_relation}

If two particles $M_1$ and $M_2$ are identical, the state of the
two-particle system must be symmetric or antisymmetric under the
interchange of two integer or half-integer spin particles.
In this subsection we work out the constraints on the covariant three-point vertex imposed by the Bose or Fermi symmetry.\s

{\it In the bosonic case with two identical particles of an integer
spin $s$,} the covariant three-point vertex tensor
$\Gamma_{\alpha\beta;\mu}$ must be symmetric under the interchange
of $M_1$ and $M_2$ due to Bose symmetry as
\begin{eqnarray}
\Gamma_{\alpha_1\cdots\alpha_s,\beta_1\cdots\beta_s;\,\mu_1\cdots\mu_J}(p,k)
\,\, = \,\,
\Gamma_{\beta_1\cdots\beta_s,\alpha_1\cdots\alpha_s;\,\mu_1\cdots\mu_J}(p,-k)\,,
\end{eqnarray}
under the transformations
\begin{eqnarray}
\alpha_i \quad \leftrightarrow \quad \beta_j
\quad  \mbox{and}\quad
k_1 \quad \leftrightarrow\quad k_2\,,
\label{eq:bosonic_interchanges}
\end{eqnarray}
for any pair of $i,j=1,\cdots, s$, leaving $p=k_1+k_2$ invariant but
changing the sign of $k$ as $k\to -k$. Combining the index and momentum
interchanges with the interchange
of the $M_1$ and $M_2$ helicities, the helicity-specific three-point
vertices transform under the Bose symmetrization as
\begin{eqnarray}
[\,{\cal H}^{J,s}_{ii [\lambda_1,\lambda_2]}\,]
   \quad \leftrightarrow \quad
   (-1)^J\, [\,{\cal H}^{J,s}_{ii [\lambda_2,\lambda_1]}\,]
   \quad \mbox{and} \quad
[\,{\cal H}^{J,s}_{hh [\lambda_1,\lambda_2]}\,]
   \quad \leftrightarrow \quad
   -(-1)^J\, [\,{\cal H}^{J,s}_{hh [\lambda_2,\lambda_1]}\,]\,,
\label{eq:bose_symmetry_three_point_vertices}
\end{eqnarray}
leading to the constraints on the helicity-specific coefficients as
\begin{eqnarray}
  c^{J,s}_{ii [\lambda_1,\lambda_2]}
\, =\, (-1)^J\, c^{J,s}_{ii [\lambda_2,\lambda_1]}
\quad \mbox{and} \quad
 c^{J,s}_{hh [\lambda_1,\lambda_2]}
\, =\, -(-1)^J\, c^{J,s}_{hh [\lambda_2,\lambda_1]}\,.
\label{eq:bose_symmetry_three_point_coefficients}
\end{eqnarray}
One observation is that the diagonal $ii$ and $hh$ elements with
$\lambda_1=\lambda_2$
vanish for odd $J$ and even $J$, respectively. Another observation
is that any spin-$1$ ($J=1$) particle cannot decay into two
identical massless spin-1 particles, as the coefficient of
the only allowed terms $c^{1,1}_{ii[\lambda,\lambda]}$ vanish,
as proven more than seventy years ago~\cite{Landau:1948kw,
Yang:1950rg}. We note that the so-called Landau-Yang
theorem is generalized to the case with any values of $J$ and
$s$~\cite{Choi:2021ewa}. \s

{\it In the fermionic case with two identical fermions of a half-integer
spin $s$}, interchanging
two identical massless fermions, i.e., taking the opposite
fermion flow line~\cite{Denner:1992vza,Denner:1992me}, we can rewrite
the helicity amplitude of the decay $X\to ff$ with a massless
fermion $M=f$ as
\begin{eqnarray}
    \tilde{\cal M}^{X\to ff}_{\sigma; \lambda_1,\lambda_2}
&=& \bar{u}_2^{\beta}(k_2,\lambda_2)\,\,
    \Gamma^{\mu}_{\beta,\alpha}(p,-k)
     \,\, v_1^{\alpha}(k_1,\lambda_1)\,\,
    \epsilon_{\mu}(p,\sigma)
    \nonumber\\
&=& v_1^{\alpha T}(k_1,\lambda_1)\,\,
     \Gamma^{\mu T}_{\beta,\alpha}(p,-k)
     \,\, \bar{u}_2^{\beta T}(k_2,\lambda_2)\,\,
    \epsilon_{\mu}(p,\sigma)\,,
\label{eq:xmm_vertex_interchanged}
\end{eqnarray}
with the superscript $T$ denoting the transpose of the matrix.
Introducing the charge-conjugation operator $C$ satisfying
$C^\dagger=C^{-1}$ and $C^T=-C$ relating the $v$ spinor to
the $u$ spinor as
\begin{eqnarray}
v^\alpha(k,\lambda) =  C\bar{u}^{\alpha T}(k, \lambda)\,,
\end{eqnarray}
with $\bar{u}=u^\dagger \gamma^0$, we can rewrite the amplitude
as
\begin{eqnarray}
    \tilde{\cal M}^{X\to ff}_{\sigma; \lambda_1,\lambda_2}
&=& - \bar{u}_1^{\beta}(k_1,\lambda_1)\,\,
     C\, \Gamma^{\mu T}_{\beta,\alpha}(p,-k)\, C^{-1}
     \,\, v_2^{\alpha}(k_2,\lambda_2)\,\,
    \epsilon_{\mu}(p,\sigma)\,.
\label{eq:xmm_vertex_interchanged_rewritten}
\end{eqnarray}
Since Fermi statistics requires $\tilde{\cal M}=- {\cal M}$, the
three-point vertex tensor must satisfy the relation
\begin{eqnarray}
  C\, \Gamma^{\mu T}_{\beta,\alpha}(p,-k)\, C^{-1}
\, =\,  \Gamma^{\mu}_{\alpha,\beta}(p,k)\,,
\label{eq:fermi_symmetry_relation}
\end{eqnarray}
which enables us to classify all the allowed terms
systematically~\cite{Kayser:1982br,Kayser:1984ge,Boudjema:1990st}.\s

The basic relation for the charge-conjugation invariance of the
Dirac equation is $C\,\gamma^T_\mu\, C^{-1} =- \gamma_\mu$ with a
unitary matrix $C$. Repeatedly using the basic relation, we can derive
\begin{eqnarray}
  \Gamma^c\, \equiv\, C\, \Gamma^T\, C^{-1}
 \,=\, \epsilon_C\,\Gamma \quad
 \mbox{with}\quad
 \epsilon_C
 = \left\{\begin{array}{lll}
        +1  & \mbox{for} & \Gamma = 1, \gamma_5, \gamma_\mu\gamma_5 \,, \\[2mm]
        -1  & \mbox{for} & \Gamma = \gamma_\mu \,,
          \end{array}\right.
\label{eq:charge_conjugation_on_gamma_structure}
\end{eqnarray}
There are no further independent terms as any other operator can be
replaced by a linear combination of $1,\, \gamma_5,\, \gamma_\mu$,
and $\gamma_\mu\gamma_5$ by use of the so-called Gordon identities,
when coupled to the $u$ and $v$ spinors.\s

Consequently, according to all the transformation properties of
covariant three-point vertex operators worked out above, the
helicity-specific three-point vertices transform under the Fermi
symmetrization as
\begin{eqnarray}
{\cal H}^{J,s}_{hi [\lambda_1,\lambda_2]}
   \quad \leftrightarrow \quad
   (-1)^J\, {\cal H}^{J,s}_{hi [\lambda_2,\lambda_1]} \quad \mbox{and} \quad
{\cal H}^{J,s}_{ih [\lambda_1,\lambda_2]}
   \quad \leftrightarrow \quad
   -(-1)^J\, {\cal H}^{J,s}_{ih [\lambda_2,\lambda_1]}\,,
\label{fermionic_IP_three_point_vertex_relation}
\end{eqnarray}
leading to the constraints on the helicity-specific coefficients as
\begin{eqnarray}
  c^{J,s}_{hi [\lambda_1,\lambda_2]}
\, =\, (-1)^J\, c^{J,s}_{hi [\lambda_2,\lambda_1]}
\quad \mbox{and} \quad
 c^{J,s}_{ih [\lambda_1,\lambda_2]}
\, =\, -(-1)^J\, c^{J,s}_{ih [\lambda_2,\lambda_1]}\,.
\label{fermionic_IP_coefficient_relation}
\end{eqnarray}
One observation is that the diagonal $hi$ and $ih$ elements with
$\lambda_1=\lambda_2$ vanish for odd $J$ and even $J$, respectively,
as in the bosonic case.\s

\subsection{Off-shell electromagnetic gauge-invariant vertices}
\label{subsec:off-shell_EM_vertices}

Due to the electromagnetic (EM) gauge invariance, any off-shell
photon couples to a conserved current. Therefore, in any time-like
photon exchange process involving the $\gamma^* M_1 M_2$ vertex,
the off-shell photon can be treated as a spin-1 particle of mass
$m_X=\sqrt{p^2}$. Moreover, the covariant three-point
$\gamma^* M_1M_2$ vertex can be cast into a manifestly
EM gauge-invariant form~\cite{Scadron:1968zz,Boudjema:1990st} as
\begin{eqnarray}
     \Gamma^\mu_{{\rm EM}\,\,
      \alpha,\beta}
\, =\,
     p^2\, \Gamma^\mu_{\alpha,\beta}
     - (\, p\cdot \Gamma_{\alpha,\beta}\,)\, p^\mu\,,
\label{eq:gauge_invariant_photon_three_point_vertex}
\end{eqnarray}
automatically satisfying the current conservation condition
$p_\mu\Gamma^\mu_{{\rm EM}\, \alpha,\beta}=
p\cdot\Gamma_{{\rm EM}\, \alpha,\beta}=0$. \s

The IP condition on the redefined EM gauge-invariant three-point vertex
is identical to that on the original three-point vertex as
the momentum $p$ is invariant under Bose or Fermi symmetry. Note that
the case with an off-shell spin-$J$ particle coupled to a conserved
tensor current can be treated in a similar manner as in the off-shell
photon case.\s

\setcounter{equation}{0}

\section{Various specific examples}
\label{sec:various_specific_examples}

First, we consider the decay of a spin-$0$ particle into two spin-$s$ particles. In this case, the restriction $|\lambda_1-\lambda_2|\leq J=0$
forces $\lambda_1=\lambda_2=\lambda$ to be satisfied
so that the three-point vertex consists of the $n\,[0,s]=2s+1$
independent terms of the form
\begin{eqnarray}
      [\, {\cal H}^{0,s}_{ii[\lambda,\lambda]}\,]
\, = \, [S^0]^{s-|\lambda|} \, [S^{\hat{\lambda}}]^{|\lambda|}\,,
\end{eqnarray}
with $\lambda$ varying from $-s$ to $s$ and
$\hat{\lambda}={\rm sign}(\lambda)$ in the bosonic case with
a non-negative integer $s$, and
\begin{eqnarray}
     [\, {\cal H}^{0,s}_{hi[\lambda,\lambda]}\,]
\, = \, [S^0]^{s-|\lambda|} [P^{\hat{\lambda}}]
        [S^{\hat{\lambda}}]^{|\lambda|-1/2}\,,
\end{eqnarray}
in the fermionic case with a positive half-integer $s$. Note that
the three-point vertices do not change in number and form even in
the IP case with $J=0$, as all the scalar composite operators,
$S^0$, $S^\pm$ and $P^\pm$, are symmetric under Bose and Fermi
symmetry transformations.  \s

Second, we consider the case with $J=1$. There exist
$n\,[1,s] = 6s+1 $ independent terms decomposed into two classes.
One class with $2s+1$ terms is when two helicities are identical,
i.e., $\lambda_1=\lambda_2$. The corresponding helicity-specific
vertex is given by
\begin{eqnarray}
       [\, {\cal H}^{1,s}_{ii[\lambda,\lambda]}\,]
\, = \, [\hat{k}]\, [S^0]^{s-|\lambda|}\,
        [S^{\hat{\lambda}}]^{|\lambda|}
\, = \, [\hat{k}]\,  [\, {\cal H}^{0,s}_{ii[\lambda,\lambda]}\,]\,,
\label{eq:bosonic_1_s_vertex_set_1}
\end{eqnarray}
in the bosonic case, and
\begin{eqnarray}
       [\, {\cal H}^{1,s}_{hi[\lambda,\lambda]}\,]
\, = \, [\hat{k}]\, [S^0]^{s-|\lambda|} [P^{\hat{\lambda}}]
        [S^{\hat{\lambda}}]^{|\lambda|-1/2}
\, = \, [\hat{k}]\, [\, {\cal H}^{0,s}_{hi[\lambda,\lambda]}\,]\,,
\label{eq:fermionic_1_s_vertex_set_1}
\end{eqnarray}
in the fermionic case. It is noteworthy that all the helicity-specific
vertices in Eqs.~(\ref{eq:bosonic_1_s_vertex_set_1}) and
(\ref{eq:fermionic_1_s_vertex_set_1}) vanish in the IP case, as
the operator $[\hat{k}]$ is antisymmetric under Bose and Fermi
symmetrization. The other class with $4s$ independent terms
is when the difference of two helicities are $\pm 1$, i.e.,
$\lambda_1=\lambda_2\pm 1$. The corresponding helicity-specific
bosonic vertex is given by
\begin{eqnarray}
       [\, {\cal H}^{1,s}_{hh[\lambda,\lambda\mp 1]}\,]
\, = \, [S^0]^{s-|\lambda_+|-1/2}\,
        [\, \delta_{\hat{\lambda}_+,\pm} V^\pm_1
        +\delta_{\hat{\lambda}_+,\mp} V^\mp_2\,]\,
        [S^{\hat{\lambda}_+}]^{|\lambda_+|-1/2}\,,
\end{eqnarray}
with $\lambda_+=\lambda\mp 1/2$ and the constraint
$|\lambda_+|\leq s-1/2$,
and the helicity-specific fermionic vertex by
\begin{eqnarray}
       [\, {\cal H}^{1,s}_{ih[\lambda,\lambda\mp 1]}\,]
\, = \, [S^0]^{s-|\lambda_+|-1/2}\,
        [W^\pm]\,
        [S^{\hat{\lambda}_+}]^{|\lambda_+|}\,,
\end{eqnarray}
with $\lambda_+=\lambda\mp 1/2$ and the constraint
$|\lambda_+|\leq s-1/2$.
In the IP case with two identical particles ($M_1=M_2$), the
covariant three-point vertex in
the bosonic case is
\begin{eqnarray}
       [\, {\cal H}^{1,s}_{hh[\lambda,\lambda\mp 1]}\,]_{\rm IP}
\, = \, \frac{1}{2}\, [S^0]^{s-|\lambda_+|-1/2}\,
        [\, \delta_{\hat{\lambda}_+,\pm} (V^\pm_1+V^\pm_2)
        +\delta_{\hat{\lambda}_+,\mp} (V^\mp_1+V^\mp_2)\,]\,
        [S^{\hat{\lambda}_+}]^{|\lambda_+|-1/2}\,,
\end{eqnarray}
with $(V^\pm_1+V^\pm_2)_{\alpha,\beta;\mu}
=(m/m_X)\,[\, (g_{\bot\alpha\mu}\hat{p}_\beta
             +g_{\bot\beta\mu}\hat{p}_\alpha)
           \pm i (\langle \alpha\mu\hat{p}\hat{k}\rangle \hat{p}_\beta
                -\langle \beta\mu\hat{p}\hat{k}\rangle \hat{p}_\alpha)\,
                ]$,
and the covariant three-point vertex in the fermionic case is
\begin{eqnarray}
       [\, {\cal H}^{1,s}_{ih[\lambda,\lambda\mp 1]}\,]_{\rm IP}
\, = \, \frac{1}{2}\, [S^0]^{s-|\lambda_+|-1/2}\,
        [W^+ + W^-]\,
        [S^{\hat{\lambda}_+}]^{|\lambda_+|} \,,
\end{eqnarray}
with $(W^+ + W^-)_\mu = \gamma_\mu \gamma_5 /\sqrt{2} m_X$ which is
of a typical axial-vector type. These results are consistent with
those in Ref.~\cite{Boudjema:1990st}. Explicitly, for $J=1$ and $s=1/2$,
the surviving three-point vertex is simply proportional to
$\gamma_\mu\gamma_5$. For $J=1$ and $s=1$, the three-point vertex,
which can be applied to the model-independent description of
the anomalous $VZZ$ vertices with virtual $V=\gamma, Z$ or
on-shell $Z'$~\cite{Hagiwara:1986vm, Gaemers:1978hg,
Gounaris:1999kf,Baur:2000ae,Keung:2008ve}, is
composed of two independent terms $(V^\pm_1+V^\pm_2)_{\alpha\beta;\mu}$
proportional to the mass $m$ so that it is vanishing in the
massless limit. \s

Third, we consider the case with $J=2$. The number of independent
terms are $n\,[2,0]=1$ for $s=0$ and $n\,[2,s]= 10s-1$ for $s>0$,
which are decomposed into two classes. The first class with $6s+1$
terms is when two helicities are identical, i.e., $\lambda_1=\lambda_2$.
The corresponding helicity-specific vertices are given by
\begin{eqnarray}
 [\, {\cal H}^{2,s}_{ii[\lambda,\lambda]}\,]
\,=\, [\hat{k}]^2\, [\,{\cal H}^{0,s}_{ii[\lambda,\lambda]}\,]
  \ \ \mbox{and}\ \
 [\, {\cal H}^{2,s}_{hh[\lambda,\lambda\pm 1]}\,]
\,=\, [\hat{k}]\, [\,{\cal H}^{1,s}_{hh[\lambda,\lambda\pm 1]}\,]\,,
\end{eqnarray}
in the bosonic case, and
\begin{eqnarray}
 [\, {\cal H}^{2,s}_{hi[\lambda,\lambda]}\,]
\,=\, [\hat{k}]^2\, [\, {\cal H}^{0,s}_{hi[\lambda,\lambda]}\,]
 \ \ \mbox{and} \ \
 [\, {\cal H}^{2,s}_{ih[\lambda,\lambda\pm 1]}\,]
\,=\, [\hat{k}]\, [\, {\cal H}^{1,s}_{ih[\lambda,\lambda\pm 1]}\,]\,,
\end{eqnarray}
in the fermionic case. They constitute $2s+1$ and $4s$ independent
covariant three-point vertices both in the bosonic and fermion
cases. The second class with $2(2s-1)$ independent terms appear for
the helicity difference of $\lambda_1-\lambda_2=\pm 2$.
The corresponding helicity-specific vertices are given by
\begin{eqnarray}
 [\, {\cal H}^{2,s}_{ii[\lambda,\lambda\mp 2]}\,]
\,=\, [S^0]^{s-|\lambda_+|-1}\,\,
      [S^{\hat{\lambda}_+}|^{|\lambda_+|}\,\,
      [T^\pm]\,,
\end{eqnarray}
in the bosonic case with $\lambda_+=\lambda\mp 1$, and
\begin{eqnarray}
 [\, {\cal H}^{2,s}_{hi[\lambda,\lambda\mp 2]}\,]
\,=\, [S^0]^{s-|\lambda_+|-1}\,\,
      [P^{\hat{\lambda}_+}]\,\,
      [S^{\hat{\lambda}_+}]^{|\lambda_+|-1/2}\,\,
      [T^\pm ]\,,
\end{eqnarray}
in the fermionic case with $\lambda_+=\lambda\mp 1$. The $J=2$ covariant
three-point vertices can be adopted for studying the massive KK graviton
or off-shell graviton interactions with two spin-$s$ particles of
equal mass $m$. \s

Now, we demonstrate the power of the algorithm by explicitly working
out all the characteristic features of four specific decay modes with the
$[J,s]$ values of $[0, 0]$, $[0,1]$, $[1,1]$ and $[2, 1]$ in the integer spin
$s$ case.  We fully write down the number of independent terms $n\,[J,s]$,
the allowed helicity assignments $(\lambda_1,\lambda_2)$, the helicity-specific covariant three-point vertices
$[\, {\cal H}^{J,s}_{[\lambda_1,\lambda_2]}\,]$ in an operator form
and their corresponding reduced helicity amplitudes ${\cal C}^J_{\lambda_1,
\lambda_2}$ as well as the helicity-specific covariant three-point vertices
$[\, {\cal H}^{J,s}_{{\rm IP} [\lambda_1,\lambda_2]}\,]$ in an operator
form and the number of independent terms $n\,[J,s]_{\rm IP}$ in the IP
case. The results are summarized succinctly
in Table~\ref{tab:specific_bosonic_examples}.\s

\vskip 0.5cm

\begin{table}[H]
\centering
\begin{tabular}{||c|c|c|c|c|c|c||}\hline\hline
\multicolumn{7}{||c||}{\color{black} \bf Integer
  spin-\boldmath{$s$} case}
  \rule{0in}{3ex} \\[2mm]
\hline\hline
  $(J,s)$
& $n\,[J,s] $
& $(\lambda_1,\lambda_2)$
& $[\, H^{J,s}_{[\lambda_1,\lambda_2]}\,]$
& ${\cal C}^J_{\lambda_1,\lambda_2}$
& $[\, H^{J,s}_{{\rm IP}[\lambda_1,\lambda_2]}\,]$
& $n\,[J,s]_{\rm IP}$ \rule{0in}{3ex} \\[2mm]
\hline
  $(0,0)$
& $1$
& $(0,0)$
& $[\,1\,]$
& $1$
& $[1]$
& $1$ \rule{0in}{3ex} \\[2mm]
\hline
  $(0,1)$
& $3$
& $(\pm 1,\pm 1)$
& $[\, S^\pm\,]$
& $1$
& $[S^\pm]$
& $3$ \rule{0in}{3ex} \\[2mm] \cline{3-6}
  { }
& { }
& $(0,0)$
& $[\, S^0\, ]$
& $-\kappa^2$
& $[S^0]$
& { } \rule{0in}{3ex} \\[2mm]
\hline
  $(1,1)$
& $7$
& $(\pm 1,\pm 1)$
& $[\, \hat{k}\, ] [S^\pm]$
& $-1$
& $-$
& $2$ \rule{0in}{3ex} \\[2mm] \cline{3-6}
  { }
& { }
& $(0,0)$
& $[\,\hat{k}\,] [S^0]$
& $\kappa^2$
& $-$
& { } \rule{0in}{3ex} \\[2mm] \cline{3-6}
  { }
& { }
& $(\pm 1,0)$
& $[V_1^\pm]$
& $\kappa$
& $\frac{1}{2}[V^\pm_1+V^\pm_2]$
& { } \rule{0in}{3ex} \\[2mm] \cline{3-5}
  { }
& { }
& $(0,\pm 1)$
& $[V_2^\pm ]$
& $-\kappa$
& { }
& { } \rule{0in}{3ex} \\[2mm]
\hline
  $(2,1)$
& $9$
& $(\pm 1,\pm 1)$
& $[\,\hat{k}\,]^2 [S^\pm]$
& $1$
& $[\,\hat{k}\,]^2\, [S^\pm]$
& $6$ \rule{0in}{3ex} \\[2mm] \cline{3-6}
  { }
& { }
& $(0,0)$
& $[\,\hat{k}\,]^2 \, [S^0] $
& $-\kappa^2$
& $[\, \hat{k}\,]^2\, [S^0]$
& { } \rule{0in}{3ex} \\[2mm] \cline{3-6}
  { }
& { }
& $(\pm 1,0)$
& $[\,\hat{k}\,] [V_1^\pm]$
& $-\kappa$
& $\frac{1}{2}\, [\,\hat{k}\,]\,
   [V^\pm_1-V^\pm_2]$
& { } \rule{0in}{3ex} \\[2mm] \cline{3-5}
  { }
& { }
& $(0,\pm 1)$
& $[\,\hat{k}\,] [V_2^\pm]$
& $\kappa$
& { }
& { } \rule{0in}{3ex} \\[2mm] \cline{3-6}
  { }
& { }
& $(\pm 1,\mp 1)$
& $[T^\pm]$
& $1$
& $\frac{1}{2}\,[T^+ + T^-] $
& { } \rule{0in}{3ex} \\[2mm]
\hline\hline
\end{tabular}
\vskip 0.3cm
\caption{\it Specific examples in the integer spin $s$ case. Listed are the number of independent terms $n\,[J,s]$, the allowed
helicity assignments $(\lambda_1,\lambda_2)$, the helicity-specific covariant
three-point vertices $[\, H^{J,s}_{[\lambda_1,\lambda_2]}\,]$ in an operator
form and their corresponding
reduced helicity amplitudes ${\cal C}^J_{\lambda_1,\lambda_2}$,
the helicity-specific covariant three-point vertex
$[\, H^{J,s}_{{\rm IP}[\lambda_1,\lambda_2]}\,]$ in an operator form and
the number of independent terms $n\,[J,s]_{\rm IP}$ in the IP case
for a few integer $J$ and integer $s$ assignments. This table is presented
for demonstrating the effectiveness of the algorithm for weaving and
characterizing the covariant triple vertices.
}
\label{tab:specific_bosonic_examples}
\end{table}

Using the algorithm for weaving the general covariant three-point
vertices by explicitly evaluating three specific decay modes with the
$[J,s]$ values of $[0,1/2]$, $[1,1/2]$ and $[2,1/2]$ in the half-integer
spin $s$ case. The results are summarized in Table~\ref{tab:specific_fermionic_examples}.\s

\vskip 0.5cm

\begin{table}[htp]
\centering
\begin{tabular}{||c|c|c|c|c|c|c||}\hline\hline
\multicolumn{7}{||c||}{\color{black} \bf Half-integer
 spin-\boldmath{$s$} case}
  \rule{0in}{3ex} \\[2mm]
\hline\hline
  $(J,s)$
& $n\,[J,s] $
& $(\lambda_1,\lambda_2)$
& $[\, H^{J,s}_{[\lambda_1,\lambda_2]}\,]$
& ${\cal C}^J_{\lambda_1,\lambda_2}$
& $[\, {\cal H}^{J,s}_{{\rm IP}[\lambda_1,\lambda_2]}\,]$
& $n\,[J,s]_{\rm IP}$ \rule{0in}{3ex} \\[2mm]
\hline
  $(0,1/2)$
& $2$
& $(\pm 1/2,\pm 1/2)$
& $P^\pm$
& $\kappa$
& $[P^\pm]$
& $2$ \rule{0in}{3ex} \\[2mm]
\hline
  $(1,1/2)$
& $4$
& $(\pm 1/2,\pm 1/2)$
& $[\,\hat{k}] [P^+]$
& $-\kappa $
& $-$
& $1$ \rule{0in}{3ex} \\[2mm] \cline{3-6}
  { }
& { }
& $(\pm 1/2, \mp 1/2)$
& $[W^\pm]$
& $\pm\kappa$
& $[W^+ + W^-]$
& { } \rule{0in}{3ex} \\[2mm]
\hline
  $(2,1/2)$
& $4$
& $(\pm 1/2,\pm 1/2)$
& $[\,\hat{k}\,]^2 [P^\pm]$
& $\kappa$
& $[\,\hat{k}\,]^2\, [P^\pm]$
& $3$ \rule{0in}{3ex} \\[2mm] \cline{3-6}
  { }
& { }
& $(\pm 1/2,\mp 1/2)$
& $[\,\hat{k}\,] [W^\pm]$
& $\mp\kappa$
& $[\,\hat{k}]\, [W^+ - W^-]$
& { } \rule{0in}{3ex} \\[2mm]
\hline\hline
\end{tabular}
\vskip 0.3cm
\caption{\it Specific examples in the half-integer spin $s$ case.
Listed are the number of independent terms $n\,[J,s]$, the allowed
helicity assignments $(\lambda_1,\lambda_2)$, the helicity-specific covariant
triple vertices $[\, H^{J,s}_{[\lambda_1,\lambda_2]}\,]$ in an operator form
and their corresponding reduced helicity amplitudes
${\cal C}^J_{\lambda_1,\lambda_2}$,
the helicity-specific covariant three-point vertex
$[\, H^{J,s}_{{\rm IP}[\lambda_1,\lambda_2]}\,]$ in an operator form and
the number of independent terms $n\,[J,s]_{\rm IP}$ in the IP case
for a few integer $J$ and half-integer $s$ assignments.
}
\label{tab:specific_fermionic_examples}
\end{table}

\setcounter{equation}{0}

\section{Conclusions}
\label{sec:conclusions}

We have developed an efficient algorithm for compactly weaving all
the covariant three-point vertices for the decay of a spin-$J$
particle $X$ of mass $m_X$ into two particles $M_{1,2}$ with equal mass
$m$ and spin $s$. For this development, we have made good use of
the closely-related equivalence between the helicity formalism and
the covariant formulation for identifying the basic building blocks
and composite three-point vertex operators for constructing all
the covariant three-point vertices. All the helicity-specific
covariant three-point vertices are presented in an operator form
in Eqs.~(\ref{eq:ii_helicity_specific_vertices})
and (\ref{eq:hh_helicity_specific_vertices}) in the bosonic case
and in Eqs.~(\ref{eq:hi_helicity_specific_vertices})
and (\ref{eq:ih_helicity_specific_vertices}) in the fermionic case,
respectively. The massless ($m=0$)
case could be worked out straightforwardly
and the (anti)symmetrization of the covariant three-point vertices
required by Bose or Fermi spin statistics of two identical
final-state particles could be made systematically in the context
of this efficient algorithm. \s

This general algorithm for constructing the effective covariant
three-point vertices is expected to be very useful in studying
various phenomenological aspects such as the indirect and direct
searches of high-spin dark matter particles and the pair production
of high-spin particles at high energy colliders.\s

Naturally, it will be valuable to extend our algorithm for dealing
with the general case when all the three particles have different
masses and spins. It is also an interesting question whether the bosonic
and fermionic cases can be synthesized in a unified framework, covering
various forms of wave tensors for particles of any spin.
These generalization and synthesis are presently under study and the
results will be reported separately.\s

\section*{Acknowledgments}
\label{sec:acknowledgments}

The work was in part by the Basic Science Research Program of Ministry of
Education through National Research Foundation of Korea
(Grant No. NRF-2016R1D1A3B01010529) and in part by the CERN-Korea theory
collaboration.\s

\end{document}